\documentclass%
[aps,prd,reprint,showpacs,longbibliography,amsfonts,amssymb,amsmath]{revtex4-1}

\usepackage[T1]{fontenc}

\usepackage{graphicx}
\usepackage{bm} % bold Greek letters
\usepackage[justification=centerlast]{caption}
\usepackage[listofformat=parens]{subfig}
\usepackage{braket} % braket notation

\makeatletter

\usepackage[
pdfauthor={Malte F. Linder, Ralf Sch\"utzhold, and William G. Unruh},
unicode=true,pdfusetitle,bookmarks=true,bookmarksnumbered=false,
bookmarksopen=true,bookmarksopenlevel=1,breaklinks=false,pdfborder={0 0 1},
backref=false,colorlinks=false]{hyperref}

\makeatother

\newcommand{\I}{\mathrm{i}} % imaginary unit
\newcommand{\E}{\mathrm{e}} % Euler's number

\newcommand{\D}{\mathrm{d}}               % differential d
\renewcommand{\Re}{\operatorname{Re}}     % real part
\renewcommand{\Im}{\operatorname{Im}}     % imaginary part
\DeclareMathOperator{\sgn}{sgn}           % signum function
\DeclareMathOperator{\Arg}{Arg}           % principal complex phase in ]-pi,pi]
\DeclareMathOperator{\Ln}{Ln}             % principal value, Im[Ln(z)] = Arg(z)
\DeclareMathOperator{\GammaFunc}{\Gamma}  % gamma function
\DeclareMathOperator{\order}{\mathcal{O}} % order

\newcommand{\vect}[1]{\bm{#1}}       % vector
\newcommand{\unitvect}[1]{\vect{#1}} % unit vector

\newcommand{\ey}{\unitvect{e}_{y}}

\newcommand{\IP}[2]{\left\langle #1,#2\right\rangle}

\newcommand{\energy}{\mathcal{E}} % energy
\newcommand{\Efield}{E}           % electric field strength

\newcommand{\cvac}{c_0}                 % vacuum speed of light
\newcommand{\epsilonvac}{\varepsilon_0} % vacuum permittivity
\newcommand{\muvac}{\mu_0}              % vacuum permeability
\newcommand{\GN}{G_{\mathrm{N}}}        % gravitational constant

\newcommand{\LagrDens}{\mathcal{L}}          % general symbol
\newcommand{\Llab}{\LagrDens_{\mathrm{lab}}} % laboratory frame
\newcommand{\Lpls}{\LagrDens_{\mathrm{pls}}} % pulse frame
\newcommand{\Leff}{\LagrDens_{\mathrm{eff}}} % low-energy eff. theory in lab.

\newcommand{\clabph}{c_{\mathrm{lab}}}          % frequency-dependent phase
\newcommand{\clow}{c_{\mathrm{low}}}            % above quantity for frequency 0
\newcommand{\vplsgr}[1][]{v_{\mathrm{gr}}^{#1}} % group velocity of a JWKB-mode
\newcommand{\JWKBktilde}{\tilde{k}_{\star}}             % general symbol
\newcommand{\JWKBktildeInd}[1]{\JWKBktilde^{#1}}        % specific mode (index)
\newcommand{\ktildestarplus}{\JWKBktildeInd{+}}         % evanescent + mode
\newcommand{\ktildestarminus}{\JWKBktildeInd{-}}        % evanescent - mode
\newcommand{\ktildestarplusminus}{\JWKBktildeInd{\pm}}  % evanescent +/- modes
\newcommand{\ktildestarp}{\JWKBktildeInd{p}}            % Hawking partner mode
\newcommand{\ktildestarcp}{\JWKBktildeInd{\mathrm{cp}}} % counterprop. mode

\newcommand{\JWKBk}{k_{\star}}                % general symbol
\newcommand{\JWKBkInd}[1]{\JWKBk^{#1}}        % specific mode with index
\newcommand{\kstarplus}{\JWKBkInd{+}}         % rapidly oscillating + mode
\newcommand{\kstarminus}{\JWKBkInd{-}}        % rapidly oscillating - mode
\newcommand{\kstarplusminus}{\JWKBkInd{\pm}}  % rapidly oscillating +/- modes
\newcommand{\kstarH}{\JWKBkInd{H}}            % Hawking mode
\newcommand{\kstarcp}{\JWKBkInd{\mathrm{cp}}} % counterpropagating mode

\newcommand{\ktildes}{\tilde{k}_{s}}           % general saddle point symbol
\newcommand{\ktildesplus}{\ktildes^{+}}        % pos.-imag.-part saddle point
\newcommand{\ktildesminus}{\ktildes^{-}}       % neg.-imag.-part saddle point
\newcommand{\ktildesplusminus}{\ktildes^{\pm}} % pos./neg.-imag.-part saddle p.
\newcommand{\ks}{k_{s}}                          % general saddle point symbol
\newcommand{\ksplus}{\ks^{+}}                    % positive saddle point
\newcommand{\ksminus}{\ks^{-}}                   % negative saddle point
\newcommand{\ksplusminus}{\ks^{\pm}}             % pos./neg. saddle points
\newcommand{\psicutp}{\psi^{\mathrm{cut}{+}}}    % pos.-real-part cut integral
\newcommand{\psicutm}{\psi^{\mathrm{cut}{-}}}    % neg.-real-part cut integral
\newcommand{\psicutpm}{\psi^{\mathrm{cut}{\pm}}} % pos./neg.-real-part integrals
\newcommand{\jcutp}{j^{\mathrm{cut}{+}}} % asymptotic Hawking mode
\newcommand{\jcutm}{j^{\mathrm{cut}{-}}} % asymptotic counterpropagating mode

\newcommand{\metric}{\mathfrak{g}}                    % general metric symbol
\newcommand{\effmetric}{\metric^{\mathrm{eff}}}       % effective metric
\newcommand{\propertime}{s}                           % gen. proper time symbol
\newcommand{\proptimeeff}{\propertime_{\mathrm{eff}}} % PT in effective metric
\newcommand{\Veff}{V_{\mathrm{eff}}}                  % effective potential

\newcommand{\aOp}{\hat{\mathfrak{a}}} % general annihilation operator
\newcommand{\aplus}{\aOp^{+}}         % positive-norm short-wavelength mode
\newcommand{\aminus}{\aOp^{-}}        % negative-norm short-wavelength mode
\newcommand{\aH}{\aOp^{H}}            % Hawking mode
\newcommand{\acp}{\aOp^{\mathrm{cp}}} % counterpropagating mode

\newcommand{\nOp}{\hat{\mathfrak{n}}} % general number operator
\newcommand{\nH}{\nOp^{H}}            % Hawking mode
\newcommand{\ncp}{\nOp^{\mathrm{cp}}} % counterpropagating mode

\newcommand{\initial}{\mathrm{in}}    % initial state subscript

\newcommand{\THawking}{T_{\mathrm{H}}} % Hawking temperature
\newcommand{\GBF}{\Gamma}              % gray-body factor

\newcommand{\IC}{\mathcal{C}}           % general symbol
\newcommand{\ICin}{\IC_{\mathrm{in}}}   % inside black hole
\newcommand{\ICout}{\IC_{\mathrm{out}}} % outside black hole

\newcommand{\TIHH}{T_{\mathrm{in}}} % temperature long-wavelength in-modes

\begin{document}

\title{Derivation of Hawking radiation in dispersive dielectric media}

\author{Malte F. \surname{Linder} and Ralf \surname{Sch\"utzhold}}
\email{ralf.schuetzhold@uni-due.de}
\affiliation{Fakult\"at~f\"ur~Physik, Universit\"at~Duisburg-Essen,
Lotharstra{\ss}e~1, 47057~Duisburg, Germany}

\author{William G. \surname{Unruh}}
\email{unruh@physics.ubc.ca}
\affiliation{Department~of~Physics~and~Astronomy,
University~of~British~Columbia,
%6224 Agricultural Road,
Vancouver, British~Columbia~V6T~1Z1, Canada}

\date{May 6, 2016}

\begin{abstract}
Motivated by recent experimental efforts, we study a black hole analog induced
by the propagation of a strong laser pulse in a nonlinear dielectric medium.
Based on the Hopfield model (one pair of Sellmeier coefficients), we perform 
an analytic and fully relativistic microscopic derivation of the analog 
of Hawking radiation in this setup.
The Hawking temperature is determined by the analog of the surface gravity (as
expected), but we also find a fre\-quen\-cy-de\-pend\-ent gray-body factor
(i.e., a nonthermal spectrum at infinity) due to the breaking of conformal
invariance in this setup.
\end{abstract}


\maketitle

%%%%%%%%%%%%%%%%%%%%%%%%%%%%%%%%%%%%%%%%%%%%%%%%%%%%%%%%%%%%%%%%%%%%%%%%%%%%%%%%
%%%%%%%%%%%%%%%%%%%%%%%%%%%%%%%%%%%%%%%%%%%%%%%%%%%%%%%%%%%%%%%%%%%%%%%%%%%%%%%%
\section{Introduction}
%%%%%%%%%%%%%%%%%%%%%%%%%%%%%%%%%%%%%%%%%%%%%%%%%%%%%%%%%%%%%%%%%%%%%%%%%%%%%%%%
%%%%%%%%%%%%%%%%%%%%%%%%%%%%%%%%%%%%%%%%%%%%%%%%%%%%%%%%%%%%%%%%%%%%%%%%%%%%%%%%

Hawking's prediction~\cite{Hawking:1,Hawking:2} that black holes evaporate due
to quantum effects has been one of the most striking consequences of quantum
field theory in curved space-times and is also expected to have profound
implications for the theory of quantum gravity.
Unfortunately, however, our chances for observing this phenomenon are very 
feeble since small enough black holes for their radiation to be observable
probably do not exist.
Nevertheless, according to the suggestion~\cite{Unruh:Sonic-analog} by one
of the authors, it might be possible to recreate this fundamental quantum effect
in the laboratory via suitable analogs.
The original proposal was based on the propagation of sound in fluids which can
generate sonic or acoustic analogs of black holes (also known as dumb holes).
The microscopic derivation of the sonic analog of Hawking radiation (including
changes in the dispersion relation at small wavelengths) has been studied by
many authors and is now quite well understood; see, e.g., Refs.~\cite{%
Unruh:Sonic-dispersion,Brout:Sonic-dispersion,Jacobson:Origin,
Corley:Sonic-dispersion,Visser:Sonic-analog,Corley:Spectrum-calculation,
Yoshiaki+Takahiro,Saida+Sakagami,Unruh+Schuetzhold:Universality,Agullo-et-al,
Schuetzhold+Unruh:Particle-origin,Macher+Parentani:BWH-radiation,
Finazzi+Parentani:Broadening,Coutant+al:S-matrix-approach,
Finazzi+Parentani:Two-regimes,Coutant+Parentani:Paradigm}.
Recently, there has been remarkable experimental progress regarding the efforts 
to observe signatures of the Hawking effect in Bose--Einstein condensates~%
\cite{BEC:Analog-1,BEC:Analog-2,BEC:Analog-3,Steinhauer:1,Steinhauer:2}.
For the sake of completeness, we would also like to mention other scenarios 
(see, e.g., Refs.~\cite{He-3-analog,Slow-light:1,Slow-light:2,Slow-light:3,
Slow-light:4,Slow-light:5,Schuetzhold+Unruh:Surface-waves-analog,
Fermi-gas-analog,Schuetzhold+Unruh:Waveguide-analog,Horstwald:Ion-ring-analog})
such as water waves~\cite{Schuetzhold+Unruh:Surface-waves-analog}, where
the classical analog of the Hawking effect has been observed recently~%
\cite{Surface-waves:Experiment-1,Surface-waves:Experiment-2}.

However, apart from the sonic analogs, there is also another very interesting 
option---optical or dielectric black hole analogs or, more generally, 
electromagnetic setups~\cite{Schuetzhold:Dielectric-analog,
Philbin:Dielectric-analog,Faccio:Dielectric-analog-1,Faccio:Dielectric-analog-2,
Finazzi+Carusotto:Kinematic,Rubino:Soliton-induced,
Finazzi:Hopfield-model-results,Petev:Moving-medium,
Belgiorno:Perturbative-photon-production,
Finazzi:Hopfield-model-WH,Belgiorno:Hopfield-model-results,
Belgiorno:Hopfield-revisited}.
In these scenarios, the fluid flow is typically replaced by the motion of 
an optical or electromagnetic pulse through the material.
Even though there have been several interesting experimental efforts~%
\cite{Belgiorno:Experiment-1,Rubino:Experiment-2,
Schuetzhold:Comment-Belgiorno-experiment,Belgiorno:Experiment-reply,
Liberati:On-Belgiorno-Experiment,Unruh+Schuetzhold:On-Belgiorno-experiment}
along this line, our theoretical understanding 
(e.g., regarding the impact of dispersion) is far less advanced than in the
case of the sonic analogs~\cite{Unruh:Sonic-dispersion,Brout:Sonic-dispersion,
Jacobson:Origin,Corley:Sonic-dispersion,Visser:Sonic-analog,
Corley:Spectrum-calculation,Yoshiaki+Takahiro,Saida+Sakagami,
Unruh+Schuetzhold:Universality,Agullo-et-al,Schuetzhold+Unruh:Particle-origin,
Macher+Parentani:BWH-radiation,Finazzi+Parentani:Broadening,
Coutant+al:S-matrix-approach,Finazzi+Parentani:Two-regimes,
Coutant+Parentani:Paradigm}.
Apart from numerical simulations (see, e.g., Refs.~\cite{Rubino:Soliton-induced,
Petev:Moving-medium}), only a very few analytical results (in analogy to the
sonic case) are available.
For example, Ref.~\cite{Belgiorno:Hopfield-model-results} presents a derivation
based on an approximation where one output channel is neglected.
However, as we shall see below, this approximation is in general 
not fully justified and gives incorrect results.
As another example, the horizon is replaced by a step-func\-tion profile in
Refs.~\cite{Finazzi:Hopfield-model-results,Finazzi:Hopfield-model-WH}.
For this simplified setup, the spectrum can also be calculated analytically.
However, because the step function formally corresponds to an infinite surface
gravity, questions such as the thermality of the spectrum and the relation
between the analog Hawking temperature and the surface gravity cannot be
addressed in this simplified setup.
Here, we contribute to filling this gap (see also Refs.~\cite{%
Finazzi:Hopfield-model-results,Linder:Diploma-thesis,
Belgiorno:Perturbative-photon-production,Finazzi:Hopfield-model-WH,
Belgiorno:Hopfield-model-results}) and provide a microscopic derivation of the
analog of Hawking radiation based on a minimal set of
as\-sump\-tions/ap\-proxi\-ma\-tions.

%%%%%%%%%%%%%%%%%%%%%%%%%%%%%%%%%%%%%%%%%%%%%%%%%%%%%%%%%%%%%%%%%%%%%%%%%%%%%%%%
%%%%%%%%%%%%%%%%%%%%%%%%%%%%%%%%%%%%%%%%%%%%%%%%%%%%%%%%%%%%%%%%%%%%%%%%%%%%%%%%
\section{\label{The Model}The Model}
%%%%%%%%%%%%%%%%%%%%%%%%%%%%%%%%%%%%%%%%%%%%%%%%%%%%%%%%%%%%%%%%%%%%%%%%%%%%%%%%
%%%%%%%%%%%%%%%%%%%%%%%%%%%%%%%%%%%%%%%%%%%%%%%%%%%%%%%%%%%%%%%%%%%%%%%%%%%%%%%%

In order to describe Hawking radiation induced by a strong, classical light
pulse in a homogeneous and transparent dielectric medium, we employ the
following microscopic model, suggested in Ref.~%
\cite{Schuetzhold:Dielectric-analog} and further developed in Refs.~%
\cite{Finazzi:Hopfield-model-results,Belgiorno:Perturbative-photon-production,
Finazzi:Hopfield-model-WH,Belgiorno:Hopfield-model-results}.
This model is closely inspired by the Hopfield model~\cite{Hopfield-model:1,
Hopfield-model:2}.
For simplicity, we assume spatial symmetry of the medium and of the pulse along
the $y$ and $z$ axes (plane symmetry).
We also restrict ourselves to one fixed polarization of the pulse 
$\vect{\Efield}_{\mathrm{pulse}}\propto\ey$ {[}(1+1)-di\-men\-sion\-al model{]}.
Additional low-inten\-si\-ty light ``on top'' of the pulse (e.g., Hawking
radiation and other quantum perturbations) with the same polarization can thus
be described by the vector potential $\vect{A}(t,x)=A(t,x)\ey$ via 
$\vect{\Efield}_{\mathrm{weak}}=\partial_{t}\vect{A}$ in temporal gauge.
This weak electromagnetic field will interact with the strong pulse via the
medium's polarizable charges, which are already excited beyond the linear
range due to the (local) intensity of the strong pulse.
The weak field will cause additional deformations of the excited states;
however, we assume that these deformations are within the linear regime 
around the polarized states influenced by the pulse alone.
Hence, each polarizable charge is assumed to interact like a scalar (one
polarization) harmonic oscillator with the weak field.
We restrict the weak field to large wavelengths compared to the molecular 
scale of the dielectric (e.g., Hawking radiation with a low temperature), 
so we can consider the dielectric in the continuum limit and not worry about
the dispersion changes created due to the finite distances between the
polarizable molecules of the medium.

This model is as follows:
there is one harmonic oscillator at each point in space, the electric dipole 
displacement being described by the scalar field $\psi(t,x)$.
The eigenfrequency $\Omega>0$ of a specific oscillator depends on the local 
classical pulse intensity, so $\Omega=\Omega(t,x)$.
This change in the local frequency of the dipoles is assumed to be the only
effect of the strong pulse in our model.
In terms of the at\-oms/mole\-cules constituting the nonlinear medium, 
this local frequency models the level spacing and hence its change 
can be understood in terms of the quadratic Stark shift 
$\Omega(t,x)\approx\Omega_0-
\alpha_{\mathrm{Stark}}\vect{\Efield}^2_{\mathrm{strong}}(t,x)$.
Note that the strong pulse could in principle also modify the dipole matrix
elements which determine the coupling between the atoms and the weak field---but
we shall largely omit this effect in our model.

Neglecting any backreaction of the weak fields $A$ and $\psi$ on the strong
pulse or the frequency changes in $\Omega$ or coupling $g$ that the strong pulse
induces, the dynamics of $A$ and $\psi$ are thus described by the Lagrangian
density
($\cvac=\epsilonvac=\muvac=\hbar=k_{B}=1$)
\begin{align}
\Llab={} &
\overbrace{
\frac{1}{2}\left(|\partial_{t}A|^{2}-|\partial_{x}A|^{2}\right)
}^{\text{weak EM field}}
+\overbrace{
\frac{1}{2}\left(|\partial_{t}\psi|^{2}-\Omega^{2}|\psi|^{2}\right)
}^{\text{medium (oscillators)}}
\nonumber \\
& {}+\underbrace{g\Re(\psi\partial_{t}A^{\ast})}_{\text{interaction}}
\label{eq:Llab}
\end{align}
in the laboratory frame (rest frame of the dielectric medium).
The Lagrangian density consists of the contributions from the free weak
electromagnetic field 
${(\vect{\Efield}_{\mathrm{weak}}^{2}-\vect{B}_{\mathrm{weak}}^{2})/2}$,
the harmonic oscillators, and the interaction between the polarization
perturbation $\psi$ and the electric field 
$\vect{\Efield}_{\mathrm{weak}}=\partial_{t}A\ey$.
The nonlinear optical influence of the strong pulse on the dielectric is encoded
in the space-time--depend\-ent eigenfrequency $\Omega(t,x)$ and potentially the
coupling constant $g(t,x)>0$, which we assume in the following are both
prescribed fields.
It is advantageous for the following analysis to generalize $A$ and $\psi$ to
complex scalar fields, so $\Llab$ has been defined accordingly.

%%%%%%%%%%%%%%%%%%%%%%%%%%%%%%%%%%%%%%%%%%%%%%%%%%%%%%%%%%%%%%%%%%%%%%%%%%%%%%%%
\subsection{Speed of light in static medium}
%%%%%%%%%%%%%%%%%%%%%%%%%%%%%%%%%%%%%%%%%%%%%%%%%%%%%%%%%%%%%%%%%%%%%%%%%%%%%%%%

Let us begin by deriving the well-known Sellmeier dispersion relation from the
model~\eqref{eq:Llab} in a static medium.
This will show us how to describe a dispersive di\-e\-lec\-tric---which then 
includes the possibility of inducing analog black hole event horizons for light
via the strong light pulse (see also Refs.~%
\cite{Finazzi:Hopfield-model-results,Finazzi:Hopfield-model-WH,
Belgiorno:Hopfield-model-results}).
By means of Hamilton's principle, $\Llab$ yields the equations of motion
\begin{subequations}
\begin{align}
(\partial_{t}^{2}-\partial_{x}^{2})A & =-\partial_{t}(g\psi)
\text{ ,}
\label{eq:EOM-lab-A}
\\
(\partial_{t}^{2}+\Omega^{2})\psi & =g\partial_{t}A
\text{ .}
\label{eq:EOM-lab-psi}
\end{align}
\end{subequations}
In order to gain a rough insight into the physics of the model, we
assume a static medium $\partial_{t}\Omega=\partial_{t}g=0$ for the moment.
Then, there are stationary solutions, each with a unique frequency 
$\omega_{\mathrm{lab}}$.
For these solutions, we can substitute 
$\partial_{t}^{2}\psi\to-\omega_{\mathrm{lab}}^{2}\psi$
and 
$\partial_{t}^{2}A\to-\omega_{\mathrm{lab}}^{2}A$
in the above equations. 
Then, solving Eq.~\eqref{eq:EOM-lab-psi} for $\psi$ and inserting the 
result back into Eq.~\eqref{eq:EOM-lab-A} leads to
\begin{equation}
\left(-\frac{\omega_{\mathrm{lab}}^{2}}{\clabph^{2}}-\partial_{x}^{2}\right)A
=
\left(\frac{1}{\clabph^{2}}\partial_{t}^{2}-\partial_{x}^{2}\right)A
=0
\end{equation}
with the fre\-quency-depend\-ent (phase) velocity of light
\begin{equation}
\clabph(\omega_{\mathrm{lab}})
=
\left(1+\frac{g^{2}}{\Omega^{2}-\omega_{\mathrm{lab}}^{2}}\right)^{-1/2}
\text{.}\label{eq:c-lab-phase-estimate}
\end{equation}
Note that we obtain the dispersion relation with only one pole at 
$\omega_{\mathrm{lab}}=\Omega$ and one Sellmeier coefficient $g$ because we
considered just one polarization field, $\psi$.
Multiple medium resonances were considered in Refs.~%
\cite{Finazzi:Hopfield-model-results,Belgiorno:Perturbative-photon-production,
Finazzi:Hopfield-model-WH,Belgiorno:Hopfield-model-results}.
This was motivated by the experiments~\cite{Belgiorno:Experiment-1,
Rubino:Experiment-2} in fused silica.
Here, we assume that the material (e.g., diamond, cf.\ Refs.~%
\cite{Faccio:BH-lasers,Finazzi:Hopfield-model-results,Petev:Moving-medium})
is well approximated by one Sellmeier pole.

The model exhibits subluminal dispersion since $\clabph$ decreases
for increasing $\omega_{\mathrm{lab}}$.
At $\omega_{\mathrm{lab}}=\Omega$ (resonance frequency of the medium), 
the speed of light formally drops to zero, which marks the breakdown of the 
model.
We will consequently restrict ourselves to lower frequencies
$|\omega_{\mathrm{lab}}|<\Omega$ 
(e.g., sufficiently low Hawking temperatures) in the remainder of this paper.
For very low frequencies, the velocity of light becomes
\begin{equation}
\clow
=
\clabph(\omega_{\mathrm{lab}}=0)
=
\left(1+\frac{g^{2}}{\Omega^{2}}\right)^{-1/2}
\text{.}
\label{eq:c-low-energy}
\end{equation}
For slowly space-time--depend\-ent $\Omega$ and $g$, the 
result~\eqref{eq:c-lab-phase-estimate} can still be approximately valid
provided that $A$ and $\psi$ oscillate very fast compared to the scales 
on which $\Omega$ and $g$ vary {[}Jeffreys--Wentzel--Kramers--Brillouin (JWKB)
approximation{]}.
Within this approximation, the strong pulse can give rise to a 
space-time--depend\-ent speed of light profile $\clow(t,x)$ in the 
dielectric medium.
However, we cannot use this JWKB approximation throughout because 
Hawking radiation (or, more generally, particle creation) is precisely
associated with a breakdown of this JWKB ap\-proxi\-ma\-tion---at least in terms
of the usual coordinates $t$ and $x$; see also
Ref.~\cite{Schuetzhold+Unruh:WKB-breakdown}.
Therefore, we have to solve the exact wave equation including the full 
dispersion relation (without neglecting any $\omega$ contributions).

%%%%%%%%%%%%%%%%%%%%%%%%%%%%%%%%%%%%%%%%%%%%%%%%%%%%%%%%%%%%%%%%%%%%%%%%%%%%%%%%
\subsection{Black holes induced by uniformly moving pulses}
%%%%%%%%%%%%%%%%%%%%%%%%%%%%%%%%%%%%%%%%%%%%%%%%%%%%%%%%%%%%%%%%%%%%%%%%%%%%%%%%

We focus on strong pulses which travel through the dielectric in the positive 
$x$ direction with a constant velocity $v\in(0,1)$ and maintain their 
shapes during the propagation.
The external fields $\Omega$ and $g$ thus only depend on the quantity $x-vt$.
In this setting, the rest frame of the pulse (pulse frame) is a second preferred
frame of reference.
Its coordinates $\tau$ and $\chi$ are connected to the laboratory frame
coordinates via the Lorentz boost
\begin{equation}
\begin{pmatrix}\tau\\
\chi
\end{pmatrix}=\gamma\begin{pmatrix}1 & -v\\
-v & 1
\end{pmatrix}\begin{pmatrix}t\\
x
\end{pmatrix}\label{eq:Lorentz-boost}
\end{equation}
with the Lorentz factor $\gamma=1/\sqrt{1-v^{2}}$.
By assumption, $\Omega$ and $g$ are independent of $\tau$ in the pulse frame,
i.e., $\partial_{\tau}\Omega=\partial_{\tau}g=0$.

%%%%%%%%%%%%%%%%%%%%%%%%%%%%%%%%%%%%%%%%%%%%%%%%%%%%%%%%%%%%%%%%%%%%%%%%%%%%%%%%
\begin{figure}
\includegraphics{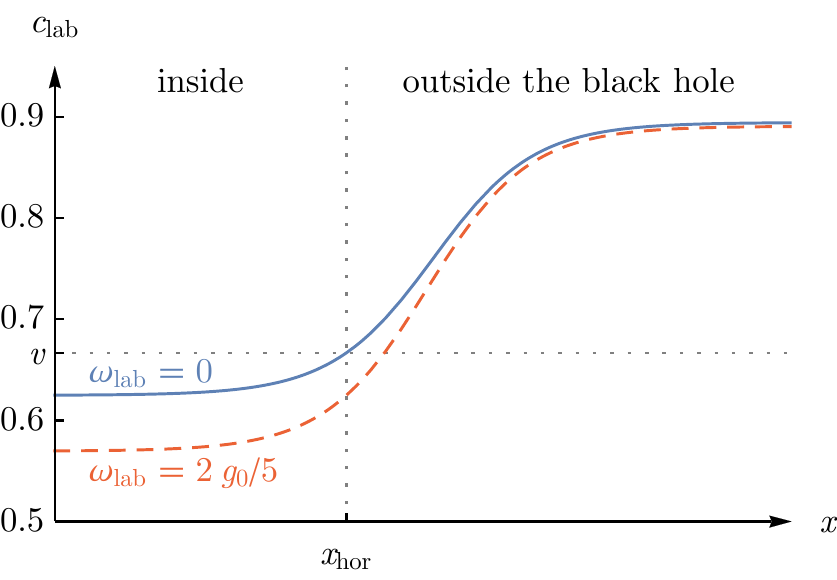}
\caption{\label{fig:pulse-profile}%
Plot of a spatial example profile of the speed of light $\clabph$ from
Eq.~\protect\eqref{eq:c-lab-phase-estimate} in the dielectric, which involves a
black hole event horizon analog.
In this example, the pulse velocity is $v=2/3$, the coupling constant
$g(t,x)=g_0$ is fixed, and the medium eigenfrequency $\Omega(t,x)$ rises from
$4g_0/5$ on the left (smaller values of $x$) to $2g_0$ on the right with a
$\tanh$ profile.
The plot shows the resulting spatial $\clabph$ curves for two different light
frequencies.
The fastest light waves occur in the low-fre\-quen\-cy limit (solid line), so
the event horizon is located at the position $x_{\mathrm{hor}}$ where these
waves propagate at the same speed as the pulse.
Light waves with higher frequencies propagate slower (subluminal dispersion) 
as indicated by the dashed line.}
\end{figure}
%%%%%%%%%%%%%%%%%%%%%%%%%%%%%%%%%%%%%%%%%%%%%%%%%%%%%%%%%%%%%%%%%%%%%%%%%%%%%%%%

A uniformly moving pulse can give rise to an analog of a black hole event
horizon in the dielectric; see, e.g.,
Refs.~\cite{Philbin:Dielectric-analog,Faccio:Dielectric-analog-1,
Faccio:Dielectric-analog-2,Finazzi:Hopfield-model-results,
Belgiorno:Perturbative-photon-production,Finazzi:Hopfield-model-WH,
Belgiorno:Hopfield-model-results}.
Considered from the laboratory point of view (moving horizon), this will happen
if the pulse has such a large intensity at its center that the weak light field
propagates slower than the pulse there ($\clow<v$), while $\clow$ exceeds $v$
outside the inner pulse region (see Fig.~\ref{fig:pulse-profile} for an example
pulse profile).
Since the dispersion is sub\-luminal---that is, light waves with low frequencies
travel at the fastest speed---the event horizon is located at the position where
the speed of light for $\omega_{\mathrm{lab}}\to0$ equals the pulse speed,
\begin{equation}
\clow=v\quad\Leftrightarrow\quad\Omega=v\gamma g
\text{ .}\label{eq:horizon-condition}
\end{equation}
A realistic pulse profile may include multiple horizons fulfilling the
condition~\eqref{eq:horizon-condition}, which could be white as well as black
hole horizons. In this paper, however, we will concentrate on a single black
hole analog event horizon and aim to calculate the corresponding Hawking
spectrum.

%%%%%%%%%%%%%%%%%%%%%%%%%%%%%%%%%%%%%%%%%%%%%%%%%%%%%%%%%%%%%%%%%%%%%%%%%%%%%%%%
%%%%%%%%%%%%%%%%%%%%%%%%%%%%%%%%%%%%%%%%%%%%%%%%%%%%%%%%%%%%%%%%%%%%%%%%%%%%%%%%
\section{Analysis in the pulse frame}
%%%%%%%%%%%%%%%%%%%%%%%%%%%%%%%%%%%%%%%%%%%%%%%%%%%%%%%%%%%%%%%%%%%%%%%%%%%%%%%%
%%%%%%%%%%%%%%%%%%%%%%%%%%%%%%%%%%%%%%%%%%%%%%%%%%%%%%%%%%%%%%%%%%%%%%%%%%%%%%%%

The pulse frame is the most advantageous frame of reference for the
derivation of the Hawking effect since the pulse and all associated
event horizons are stationary with respect to that frame.
The Lagrangian density $\Llab$ in Eq.~\eqref{eq:Llab} transformed to the pulse
frame reads
\begin{align}
\Lpls={} &
\frac{1}{2}\left(|\partial_{\tau}A|^{2}-|\partial_{\chi}A|^{2}\right)
\nonumber \\
&{}+\frac{1}{2}\left[\gamma^{2}|(\partial_{\tau}-v\partial_{\chi})\psi|^{2}
-\Omega^{2}|\psi|^{2}\right]
\nonumber \\
&{}+\gamma g\Re[\psi(\partial_{\tau}-v\partial_{\chi})A^{\ast}]
\text{ .}
\label{eq:Lpls}
\end{align}
In this section, we will combine the two equations of motion for $A$ and 
$\psi$ into a single equality and derive two important and well-known 
conservation laws from the symmetries of $\Lpls$.

%%%%%%%%%%%%%%%%%%%%%%%%%%%%%%%%%%%%%%%%%%%%%%%%%%%%%%%%%%%%%%%%%%%%%%%%%%%%%%%%
\subsection{Stationary modes}
%%%%%%%%%%%%%%%%%%%%%%%%%%%%%%%%%%%%%%%%%%%%%%%%%%%%%%%%%%%%%%%%%%%%%%%%%%%%%%%%

The time invariance of $\Lpls$ due to $\partial_{\tau}\Omega=\partial_{\tau}g=0$
implies the conservation of the frequency (energy) of any solution $(A,\psi)$ of
the equations of motion.
Hence, we may concentrate on solutions of the form
\begin{equation}
A(\tau,\chi)=A_{\omega}(\chi)\E^{-\I\omega\tau}\text{ ,}\quad
\psi(\tau,\chi)=\psi_{\omega}(\chi)\E^{-\I\omega\tau}
\label{eq:stationary-mode-separation-ansatz}
\end{equation}
with a unique, conserved pulse frame frequency $\omega$ (stationary modes).
Inserting this form into the equations of motion~\eqref{eq:EOM-lab-A}
and~\eqref{eq:EOM-lab-psi} transformed to the pulse frame
(with $\partial_{\tau}\to-\I\omega$) leads to the mode equations
\begin{subequations}
\begin{align}
-(\omega^{2}+\partial_{\chi}^{2})A_{\omega} & =
\gamma\,(\I\omega+v\partial_{\chi})g\psi_{\omega}
\text{ ,}\label{eq:mode-eq-pls-A}\\
\left[\gamma^{2}(\I\omega+v\partial_\chi)^{2}+\Omega^{2}\right]\!\psi_\omega & =
-\gamma g\,(\I\omega+v\partial_{\chi})A_{\omega}
\text{ ,}\label{eq:mode-eq-pls-psi}
\end{align}
\end{subequations}
which are satisfied by the mode functions $A_{\omega}$ and $\psi_{\omega}$.

The mode equations~\eqref{eq:mode-eq-pls-A} and~\eqref{eq:mode-eq-pls-psi}
can be combined into one single ordinary differential equation.
To this end, we apply the operator $\gamma\,(\I\omega+v\partial_{\chi})$,
which commutes with $\omega^{2}+\partial_{\chi}^{2}$, to the upper
equation~\eqref{eq:mode-eq-pls-A} and then insert Eq.~\eqref{eq:mode-eq-pls-psi}
divided by $g$.
This results in the decoupled fourth-order mode equation
\begin{multline}
\left\{ 
(\omega^{2}+\partial_{\chi}^{2})\frac{1}{g}
\left[\gamma^{2}(\I\omega+v\partial_{\chi})^{2}+\Omega^{2}\right]
\vphantom{-\gamma^{2}(\I\omega+v\partial_{\chi})^{2}g}
\right.\\
\left.\vphantom{(\omega^{2}+\partial_{\chi}^{2})
\frac{1}{g}\left[\gamma^{2}(\I\omega+v\partial_{\chi})^{2}+\Omega^{2}\right]}
-\gamma^{2}(\I\omega+v\partial_{\chi})^{2}g\right\} \psi_{\omega}=0
\label{eq:decoupled-mode-eq}
\end{multline}
for the mode function $\psi_{\omega}$. It is important to note that a difference
between this equation and the usual equations used in much of the analog model
literature is that this is not a second order equation in time.
In order to calculate the corresponding function $A_{\omega}$, which is uniquely
determined by a given solution $\psi_{\omega}$ of
Eq.~\eqref{eq:decoupled-mode-eq}, we apply the operator
$(\I\omega-v\partial_{\chi})$ on Eq.~\eqref{eq:mode-eq-pls-psi} and use this
result to eliminate the term $\partial_{\chi}^{2}A_{\omega}$ in 
Eq.~\eqref{eq:mode-eq-pls-A}, which gives
\begin{multline}
A_{\omega}
=
\frac{\gamma^{2}}{\omega^{2}}
\left\{ 
(\I\omega-v\partial_{\chi})\frac{1}{\gamma g}
\left[\gamma^{2}(\I\omega+v\partial_{\chi})^{2}+\Omega^{2}\right]
\vphantom{+v^{2}\gamma\,(\I\omega+v\partial_{\chi})g}
\right.\\
\left.\vphantom{(\I\omega-v\partial_{\chi})\frac{1}{\gamma g}
\left[\gamma^{2}(\I\omega+v\partial_{\chi})^{2}+\Omega^{2}\right]}
+v^{2}\gamma\,(\I\omega+v\partial_{\chi})g\right\} \psi_{\omega}
\text{ .}\label{eq:mode-func-A-from-psi}
\end{multline}
Note that one should be careful with the order in the equations above as we are
considering nonhomogeneous, i.e., $\chi$-depend\-ent profiles, such that
$(\I\omega-v\partial_{\chi})$ does not commute with $\Omega$, for example.

%%%%%%%%%%%%%%%%%%%%%%%%%%%%%%%%%%%%%%%%%%%%%%%%%%%%%%%%%%%%%%%%%%%%%%%%%%%%%%%%
\subsection{\label{sub:Noether-charge-conservation}Conserved generalized inner
product and norm}
%%%%%%%%%%%%%%%%%%%%%%%%%%%%%%%%%%%%%%%%%%%%%%%%%%%%%%%%%%%%%%%%%%%%%%%%%%%%%%%%

Since we have generalized our physical model to complex fields $A$ and $\psi$,
the Lagrangian density $\Lpls$ is invariant under any transformation of the
global phase of the dynamic fields ($A\to\E^{\I\varphi}A$ and
$\psi\to\E^{\I\varphi}\psi$).
By means of Noether's theorem, this continuous symmetry of $\Lpls$ is related 
to a conserved current $\partial_{\tau}\rho+\partial_{\chi}j=0$ with the
(Noether) charge density (see also
Refs.~\cite{Finazzi:Hopfield-model-results,Belgiorno:Hopfield-model-results})
\begin{align}
\rho 
& =
\I
\left(A^{\ast}\Pi_{A^{\ast}}+\psi^{\ast}\Pi_{\psi^{\ast}}
-\Pi_{A}A-\Pi_{\psi}\psi\right)
\nonumber \\
& =
-\Im\!\left[
A^{\ast}\,(\partial_{\tau}A+\gamma g\psi)
+\gamma^{2}\psi^{\ast}\,(\partial_{\tau}-v\partial_{\chi})\psi
\right]
\label{eq:Noether-charge-density}
\end{align}
and the current density
\begin{equation}
j=\Im\!\left[
A^{\ast}\,(\partial_{\chi}A+v\gamma g\psi)
+v\gamma^{2}\psi^{\ast}\,(\partial_{\tau}-v\partial_{\chi})\psi
\right]
\text{.}\label{eq:Noether-current-density}
\end{equation}
The canonical momentum densities appearing in $\rho$ are given by
\begin{alignat}{3}
\Pi_{A} & =
\frac{\partial\Lpls}{\partial(\partial_{\tau}A)} 
&  & =\frac{1}{2}(\partial_{\tau}A^{\ast}+\gamma g\psi^{\ast}) 
&  & =\left(\Pi_{A^{\ast}}\right)^{\ast}
\text{,}\\
\Pi_{\psi} & =
\frac{\partial\Lpls}{\partial(\partial_{\tau}\psi)} 
&  & =\frac{\gamma^{2}}{2}(\partial_{\tau}-v\partial_{\chi})\psi^{\ast} 
&  & =\left(\Pi_{\psi^{\ast}}\right)^{\ast}
\text{.}
\end{alignat}
For stationary modes of the form~\eqref{eq:stationary-mode-separation-ansatz},
all time dependencies in $\rho$ and $j$ cancel each other out
($\partial_{\tau}\to-\I\omega$); that is, these quantities are
time independent.
The continuity equation consequently simplifies to $\partial_{\chi}j=0$, so the
current density is a space-time--inde\-pend\-ent quantity for stationary modes.

The conservation of the total (integrated) Noether charge can be used to derive
the conserved, generalized inner product
\begin{align}
\IP{\begin{pmatrix}A_1\\ \psi_1\end{pmatrix}}
   {\begin{pmatrix}A_2\\ \psi_2\end{pmatrix}}=\I\intop_{-\infty}^{\infty} &
\left(A_1^{\ast}\Pi_{A_2^{\ast}}+\psi_1^{\ast}\Pi_{\psi_2^{\ast}}
\vphantom{-\Pi_{A_1}A_2-\Pi_{\psi_1}\psi_2}\right.
\nonumber \\
&\left.\vphantom{A_1^{\ast}\Pi_{A_2^{\ast}}+\psi_1^{\ast}\Pi_{\psi_2^{\ast}}}
-\Pi_{A_1}A_2-\Pi_{\psi_1}\psi_2\right)
\D\chi\text{ ,}
\label{eq:inner-product}
\end{align}
which is also known as the Klein--Gordon inner product~\cite{Wald:GR}, between
two arbitrary solutions, $(A_1,\psi_1)$ and $(A_2,\psi_2)$, of the equations of
motion in the pulse frame.
It has the same properties as usual inner products except for 
positive definiteness since the product of a field solution 
$(A,\psi)$ with itself coincides with its total Noether charge,
\begin{equation}
\IP{\begin{pmatrix}A\\
\psi
\end{pmatrix}}{\begin{pmatrix}A\\
\psi
\end{pmatrix}}=\intop_{-\infty}^{\infty}\rho\,\D\chi\text{ ,}\label{eq:norm}
\end{equation}
which is not necessarily positive but can be any real number.
In fact, taking the complex conjugate of Eq.~\eqref{eq:inner-product} 
shows that the inner product of $A^{\ast}$ and $\psi^{\ast}$ with itself 
has a sign opposite to that of $A$ and $\psi$.
We call this quantity~\eqref{eq:norm} the (pseudo-)norm of the field solution
$(A,\psi)$.

%%%%%%%%%%%%%%%%%%%%%%%%%%%%%%%%%%%%%%%%%%%%%%%%%%%%%%%%%%%%%%%%%%%%%%%%%%%%%%%%
%%%%%%%%%%%%%%%%%%%%%%%%%%%%%%%%%%%%%%%%%%%%%%%%%%%%%%%%%%%%%%%%%%%%%%%%%%%%%%%%
\section{JWKB analysis}
%%%%%%%%%%%%%%%%%%%%%%%%%%%%%%%%%%%%%%%%%%%%%%%%%%%%%%%%%%%%%%%%%%%%%%%%%%%%%%%%
%%%%%%%%%%%%%%%%%%%%%%%%%%%%%%%%%%%%%%%%%%%%%%%%%%%%%%%%%%%%%%%%%%%%%%%%%%%%%%%%

As a first approach to understanding the structure of the solutions of the
decoupled mode equation~\eqref{eq:decoupled-mode-eq} in the pulse frame well
outside and inside a black hole, we apply the JWKB approximation.
That is, we treat the external fields $\Omega$ and $g$ as constant
($\partial_{\chi}\Omega=\partial_{\chi}g=0$), so the mode functions
$\psi_{\omega}$ are (superpositions of) plane waves, and thus we
may set $\partial_{\chi}\to\I k$ in Eq.~\eqref{eq:decoupled-mode-eq}.
The following analysis is thus equivalent to the one in
Refs.~\cite{Finazzi:Hopfield-model-results,Finazzi:Hopfield-model-WH},
where a piecewise homogeneous setup is considered.
The resulting dispersion relation,
\begin{equation}
(\omega^{2}-k^{2})\left[\Omega^{2}-\gamma^{2}(\omega+vk)^{2}\right]
+\gamma^{2}g^{2}(\omega+vk)^{2}
=0\text{ ,}
\label{eq:disp-rel-polynomial}
\end{equation}
which is just the relation~\eqref{eq:c-lab-phase-estimate} 
transformed to the pulse frame, can be rearranged into the form
\begin{equation}
\underbrace{\gamma\left(\omega+vk\right)}_{\omega_{\mathrm{lab}}}
=
\pm F(k)
\label{eq:disp-rel-lab-frequency}
\end{equation}
with the (phase velocity) function
\begin{equation}
F(k)=\Omega\sqrt{1+\frac{g^{2}}{\omega^{2}-k^{2}-g^{2}}}
\text{ .}\label{eq:F-definition}
\end{equation}
Each solution $\JWKBk$ of Eqs.~\eqref{eq:disp-rel-polynomial} and~%
\eqref{eq:disp-rel-lab-frequency} is an allowed wave vector at the
frequency $\omega$.
To create a black hole analog as in Fig.~\ref{fig:pulse-profile},
however, $\Omega$ and/or $g$ must be inhomogeneous.
Nevertheless, the (now $\chi$-de\-pend\-ent) wave vector solutions of the 
(local) dispersion relation will still approximate the physics of the fields
well as long as the length scales on which $\Omega$ and $g$ vary
are much greater than the inverse wave vectors $1/\JWKBk$.

The dispersion relation~\eqref{eq:disp-rel-polynomial} has up to
four different and possibly complex solutions $\JWKBk$ for given
values of $v$, $\Omega$, $g$, and $\omega$.
We are particularly interested in the real solutions, which describe 
propagating waves.
The dispersion relation (as a fourth-order polynomial) can be solved 
analytically, but the resulting expressions are quite lengthy in general,
such that it is hard to grasp their physical properties. 
Therefore, we will use the form in Eq.~\eqref{eq:disp-rel-lab-frequency} 
to find the solutions graphically instead by plotting both sides of this 
equation over $k$.
The left-hand side yields a straight line.
Note that, according to the Lorentz boost in Eq.~\eqref{eq:Lorentz-boost}, 
$\gamma\,(\omega+vk)$ is the laboratory frame frequency $\omega_{\mathrm{lab}}$
of a wave which has the frequency $\omega$ and the wave vector $k$ in the
pulse frame.
For a small $|\omega|<g$, the function $F(k)\geq0$ appearing on the right-hand
side of Eq.~\eqref{eq:disp-rel-lab-frequency} is only real for $|k|\geq|\omega|$
and approaches the asymptotic value $\Omega$ for $k\to\pm\infty$.
At high pulse frame frequencies $|\omega|>g$, on the other hand, 
$F$ is also real for $|k|<\sqrt{\omega^{2}-g^{2}}$.
Solutions $\JWKBk$ in this $k$ range
(intersection points with the straight line $\omega_{\mathrm{lab}}$),
however, correspond to waves with large laboratory frame frequencies
$|\omega_{\mathrm{lab}}|>\Omega$ beyond the range of validity of
the physical model.
Hence, we restrict the analysis to small pulse frame frequencies 
with $|\omega|<g$ (or even $|\omega|\ll g$).

%%%%%%%%%%%%%%%%%%%%%%%%%%%%%%%%%%%%%%%%%%%%%%%%%%%%%%%%%%%%%%%%%%%%%%%%%%%%%%%%
\begin{figure}
\subfloat[\label{fig:disp-rel-vary-med-res}Solution for varying $\Omega$.]%
{\includegraphics{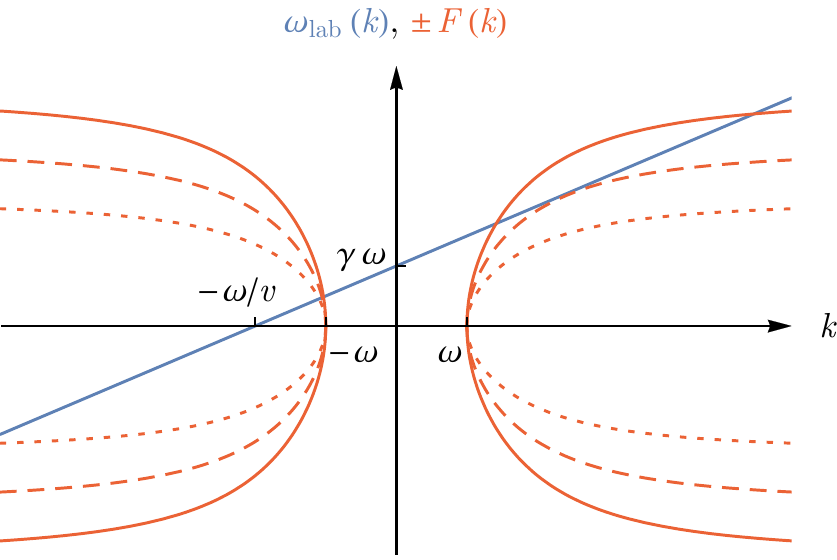}}

\subfloat[\label{fig:disp-rel-vary-mode-freq}Solution for varying $\omega$.]%
{\includegraphics{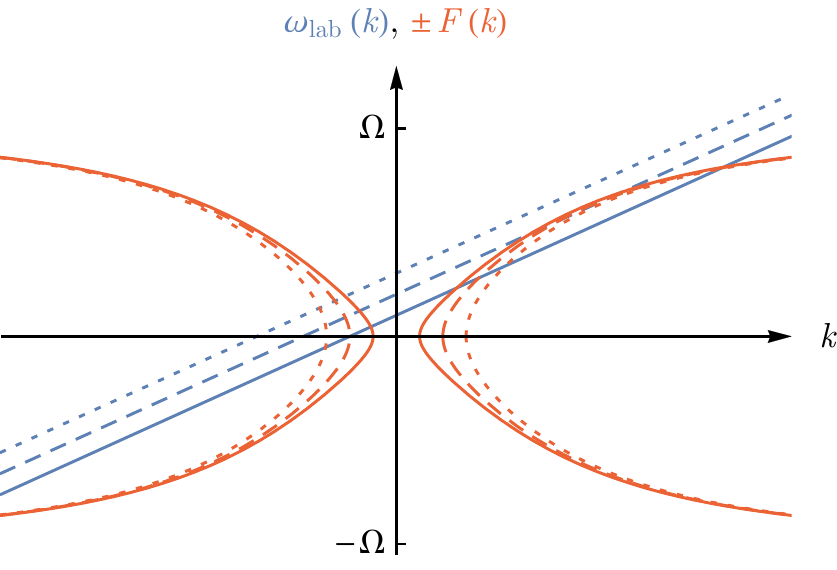}}

\caption{\label{fig:disp-rel-solution}%
Graphical solution scheme of the dispersion relation~\protect%
\eqref{eq:disp-rel-lab-frequency} for $v=1/2$ and constant $g(\chi)=g_0$.
In both plots, the straight lines are the graphs of the laboratory frame
frequency $\gamma\,(\omega+vk)$, and the curves are the graphs of $\pm F(k)$.
(\protect\subref*{fig:disp-rel-vary-med-res})~%
Solution for the fixed mode frequency $\omega=g_0/2$ and decreasing $\Omega$
(solid $\to$ dashed $\to$ dotted $\pm F$ curves).
(\protect\subref*{fig:disp-rel-vary-mode-freq})~%
Solution for constant $\Omega$ and increasing $\omega$ (solid $\to$ dashed $\to$
dotted $\omega_{\mathrm{lab}}$ lines and $\pm F$ curves).%
}
\end{figure}
%%%%%%%%%%%%%%%%%%%%%%%%%%%%%%%%%%%%%%%%%%%%%%%%%%%%%%%%%%%%%%%%%%%%%%%%%%%%%%%%

Examples for the graphical solution of the dispersion relation (including the
effects of varying parameters $\omega$ and $\Omega$) are depicted in
Fig.~\ref{fig:disp-rel-solution}.

%%%%%%%%%%%%%%%%%%%%%%%%%%%%%%%%%%%%%%%%%%%%%%%%%%%%%%%%%%%%%%%%%%%%%%%%%%%%%%%%
\subsection{\label{sub:JWKB-modes-outside}Modes outside the black hole}
%%%%%%%%%%%%%%%%%%%%%%%%%%%%%%%%%%%%%%%%%%%%%%%%%%%%%%%%%%%%%%%%%%%%%%%%%%%%%%%%

\begin{figure}
\includegraphics{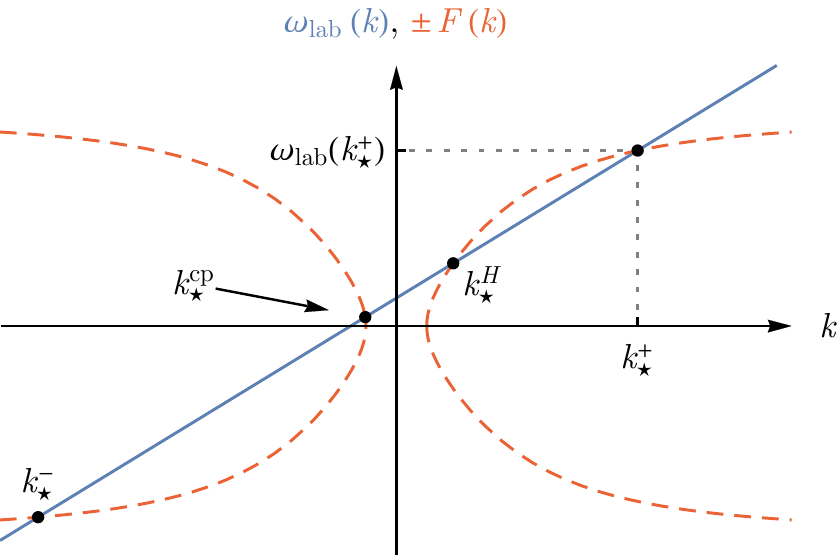}
\caption{\label{fig:disp-rel-outside}%
Graphical solution of the dispersion
relation~\protect\eqref{eq:disp-rel-lab-frequency} far outside the black
hole in Fig.~\protect\ref{fig:pulse-profile}, where $v=2/3$, $g(\chi)=g_0$
is constant, and the local value of $\Omega$ is $2g_0$. 
We consider the mode frequency $\omega=g_0/5$. 
The straight line is the laboratory frame frequency $\gamma\,(\omega+vk)$. 
The dashed lines are the graphs of $\pm F(k)$. 
Here, we get four real solutions (labeled intersection points): 
the Hawking mode $\kstarH$, the counterpropagating mode $\kstarcp$,
and the two short-wave\-length modes $\kstarplusminus$. 
The vertical value of an intersection point coincides with the 
frequency of the corresponding wave as measured in the laboratory frame.}
\end{figure}

We start to discuss the solutions of the dispersion relation far outside
an analog black hole horizon for a low frequency mode $0<\omega\ll\Omega$.
The graphical solution is depicted in Fig.~\ref{fig:disp-rel-outside}.
There are four different real solutions: two small wave vectors
$\kstarH\gtrsim\omega$ and $\kstarcp\lesssim-\omega$ as well as two large
solutions $\kstarplus\gg\omega$ and $\kstarminus\ll-\omega$. 
All four possible wave vectors outside the black hole are thus real 
(propagating modes). 
The long-wave\-length modes $\kstarH$ and $\kstarcp$ with wave numbers
of the order $\order(\omega)$ are hardly affected by dispersion, whereas 
the rapidly oscillating modes $\kstarplusminus$ are a consequence of 
dispersion (they vanish if dispersion is neglected).
Let us derive some properties of the modes.

%%%%%%%%%%%%%%%%%%%%%%%%%%%%%%%%%%%%%%%%%%%%%%%%%%%%%%%%%%%%%%%%%%%%%%%%%%%%%%%%
%%%%%%%%%%%%%%%%%%%%%%%%%%%%%%%%%%%%%%%%%%%%%%%%%%%%%%%%%%%%%%%%%%%%%%%%%%%%%%%%

By differentiating the dispersion relation~\eqref{eq:disp-rel-polynomial}
with respect to $k$ (treating $\omega$ as a function of $k$ for the moment),
we find the expression
\begin{equation}
\vplsgr[\JWKBk]
=
\left.\frac{\D\omega(k)}{\D k}\right|_{k=\JWKBk}
=
\frac{\partial_{k}[\pm F(k)]|_{k=\JWKBk}-v\gamma}
{\gamma+\partial_{k}[\pm F(k)]|_{k=\JWKBk}\omega/\JWKBk}
\label{eq:group-velocity}
\end{equation}
for the pulse frame group velocity of the mode $\JWKBk$. 
The $\pm$ sign before $F$ depends on whether $\JWKBk$ is a plus or minus 
solution of Eq.~\eqref{eq:disp-rel-lab-frequency}, i.e., on the sign of the
laboratory frame frequency of the mode (vertical coordinate of the
corresponding intersection point in Fig.~\ref{fig:disp-rel-outside}).
This formula allows us to determine the group velocity signs for all
four modes outside the black hole just based on the graphical solution
of the dispersion relation in Fig.~\ref{fig:disp-rel-outside}. 
We find that the group velocities of the modes $\kstarplus$, $\kstarminus$,
and $\kstarcp$ are negative, so these modes propagate towards the
(stationary) black hole event horizon.
The mode $\kstarcp$, which is a low-ener\-gy mode 
(i.e., almost unaffected by dispersion), corresponds to 
light propagating in the opposite direction as the pulse when viewed 
from the laboratory frame---that is, this is the counterpropagating mode.
(In the sonic black hole analogs based on flowing fluids, this would 
be the downstream mode.)
In the pulse frame, this mode moves towards the black hole and 
can cross the horizon without being distorted drastically.
However, it can also be scattered into an copropagating mode (see
Sec.~\ref{Conclusions} below), but this process is purely 
classical scattering and does not lead to particle creation.
We thus do not expect this mode to be the origin of Hawking radiation.
Consequently, we expect $\kstarplusminus$ to be the relevant in-modes
for creating the outgoing Hawking radiation as usual in systems with 
subluminal dispersion (cf., e.g., Refs.~\cite{Corley:Sonic-dispersion,
Unruh+Schuetzhold:Universality}).
The mode $\kstarH$ is the only mode with a positive group velocity, 
so these waves escape the black hole (e.g., Hawking radiation).

As a next step, we consider the Noether charge densities $\rho$
{[}see Eq.~\eqref{eq:Noether-charge-density}{]} of the modes. 
The mode function $A_{\omega}$ corresponding to a JWKB solution 
of the (approximate) form 
$\psi_{\omega}(\chi)=\exp(\I\JWKBk\chi-\I\omega\tau)$
is calculated using Eq.~\eqref{eq:mode-func-A-from-psi}, with 
$\partial_{\tau}\to-\I\omega$ (stationary modes) and 
$\partial_{\chi}\to\I\JWKBk$ (JWKB approximation).
The resulting charge density of the mode reads
\begin{equation}
\rho_{\omega}^{\JWKBk}=
\gamma^{2}(\omega+v\JWKBk)
\left[1+\frac{g^{2}\JWKBk\,(v\omega+\JWKBk)}{(\JWKBk^{2}-\omega^2)^2}\right]
\text{.}\label{eq:JWKB-mode-charge-density}
\end{equation}
The sign of this charge density depends only on the laboratory frame
frequency of the mode,
\begin{equation}
\sgn\rho_{\omega}^{\JWKBk}=\sgn(\omega+v\JWKBk)\text{ .}
\label{Noether-charge-sign}
\end{equation}
Hence, it follows from Fig.~\ref{fig:disp-rel-outside} that $\kstarminus$ is the
only mode which propagates a negative Noether charge, so the contribution from
this in-mode is the essential ingredient of pair production (cf., e.g., Refs.~%
\cite{Brout:Sonic-dispersion,Corley:Sonic-dispersion,
Corley:Spectrum-calculation,Saida+Sakagami,Unruh+Schuetzhold:Universality,
Macher+Parentani:BWH-radiation,Coutant+al:S-matrix-approach,
Finazzi+Parentani:Two-regimes,Finazzi:Hopfield-model-results,
Belgiorno:Hopfield-model-results}).

%%%%%%%%%%%%%%%%%%%%%%%%%%%%%%%%%%%%%%%%%%%%%%%%%%%%%%%%%%%%%%%%%%%%%%%%%%%%%%%%
\subsection{\label{sub:JWKB-modes-inside}Modes inside the black hole}
%%%%%%%%%%%%%%%%%%%%%%%%%%%%%%%%%%%%%%%%%%%%%%%%%%%%%%%%%%%%%%%%%%%%%%%%%%%%%%%%

\begin{figure}
\includegraphics{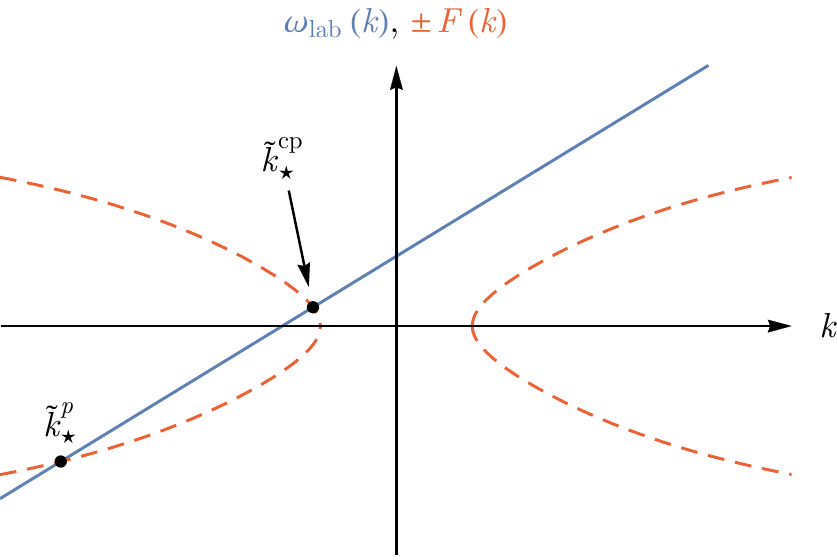}
\caption{\label{fig:disp-rel-inside}%
Graphical solution of the dispersion relation
far inside the black hole in Fig.~\protect\ref{fig:pulse-profile}. We consider
the same configuration as in Fig.~\protect\ref{fig:disp-rel-outside} except
for the value of $\Omega$, which has decreased to $4g_0/5$. The
dashed graphs of $\pm F(k)$ have ``narrowed'' correspondingly,
so there are only two real solutions $\ktildestarcp$ (counterpropagating mode)
and $\ktildestarp$ (infalling partner) inside the black hole. The two
remaining wave vector solutions $\ktildestarplusminus$ are complex
(evanescent modes).}
\end{figure}

As one goes into the black hole in Fig.~\ref{fig:pulse-profile},
$\Omega$ decreases, so the graphs of $\pm F(k)$ in the graphical
solution in Fig.~\ref{fig:disp-rel-outside} ``narrow'' because
$F(k)\propto\Omega$; cf.\ Fig.~\subref{fig:disp-rel-vary-med-res}.
The solutions $\kstarH$ and $\kstarplus$ will thus approach each other 
and merge at a certain point on the way towards the horizon, at which 
point the straight line $\gamma\,(\omega+vk)$ is tangent to $F(k)$---see the
dashed $\pm F(k)$ curves in Fig.~\subref{fig:disp-rel-vary-med-res}. 
Beyond this point, these two real solutions become two complex 
solutions, $\ktildestarplus$ and $\ktildestarminus=(\ktildestarplus)^{\ast}$,
with $\Im\ktildestarplus>0$, so the Hawking mode escaping the black
hole disappears beyond this ``point of no return'', which behaves
like a fre\-quen\-cy-spe\-cif\-ic event horizon.
The allowed wave vectors vary rapidly around this point, so the JWKB 
approximation breaks down there. 
Note that, in the limit $\omega\to0$, the point of no return coincides 
with the (``absolute'') event horizon since low-fre\-quen\-cy waves 
travel at the fastest speed (subluminal dispersion).
Deep inside the black hole in Fig.~\ref{fig:pulse-profile}, the
JWKB approximation is valid again. 
See Fig.~\ref{fig:disp-rel-inside} for the graphical solution of the 
local dispersion relation. 
Both real solutions, $\ktildestarcp$ and $\ktildestarp$, describe modes
with negative group velocities according to Eq.~\eqref{eq:group-velocity},
so they propagate deeper into the black hole (as expected inside the event
horizon).
The mode $\ktildestarcp$ has a positive Noether charge density while 
$\ktildestarp$ carries negative charge. 
We conclude that $\ktildestarcp\approx\kstarcp$ is the counterpropagating 
mode, which crossed the horizon and is now inside the black hole.
The second mode, $\ktildestarp$, with a negative laboratory frame frequency 
(negative energy) is the infalling partner mode of the outgoing 
Hawking mode $\kstarH$; see, e.g., Refs.~\cite{Hawking:2,
Brout:Sonic-dispersion}.
The mode structure inside the black hole is thus as expected according to
Refs.~\cite{Brout:Sonic-dispersion,Corley:Sonic-dispersion,
Corley:Spectrum-calculation,Saida+Sakagami,Unruh+Schuetzhold:Universality,
Macher+Parentani:BWH-radiation,Coutant+al:S-matrix-approach,
Finazzi+Parentani:Two-regimes,Finazzi:Hopfield-model-results,
Belgiorno:Hopfield-model-results} again.

%%%%%%%%%%%%%%%%%%%%%%%%%%%%%%%%%%%%%%%%%%%%%%%%%%%%%%%%%%%%%%%%%%%%%%%%%%%%%%%%
%%%%%%%%%%%%%%%%%%%%%%%%%%%%%%%%%%%%%%%%%%%%%%%%%%%%%%%%%%%%%%%%%%%%%%%%%%%%%%%%

One of the standard ways to derive the Hawking effect is to trace
a late-time outgoing Hawking wave packet back in time in order to
find the early-time ingoing wave packets the Hawking packet is 
composed of initially (see, e.g., Ref.~\cite{Corley:Sonic-dispersion}).
Since there are no solutions inside the black hole which can approach 
the horizon (subluminal dispersion), the relevant boundary condition 
for that derivation of Hawking radiation is that the mode function 
vanishes deep inside the black hole (i.e., the mode is evanescent).
In the limit of wave packets which are sharply peaked around
a unique frequency $\omega$, the wave packets become very extended
in space and time---that is, we consider single stationary modes.
We have to make sure that the mode function $\psi_{\omega}$ satisfies
the boundary condition; however, a function $\psi_{\omega}$ which
is nonzero outside the black hole cannot vanish identically 
everywhere inside the horizon~\cite{Corley:Sonic-dispersion}.
We therefore demand $\psi_{\omega}$ to decay rapidly (exponentially) 
inside the black hole. 
In terms of the JWKB solutions explained above, that means that only 
the complex wave vector $\ktildestarminus$ may contribute to 
$\psi_{\omega}$ beyond the horizon since 
$|\exp(\I\ktildestarminus\chi-\I\omega\tau)|=\exp(-\chi\Im\ktildestarminus)$
and $\Im\ktildestarminus<0$, so this function decays rapidly along
the way deeper into the black hole (decreasing $\chi$).
However, we cannot continue this mode function across the event horizon 
using the JWKB technique because this approximation breaks down in the 
vicinity of the horizon (at least in the coordinates we use here; see 
also Ref.~\cite{Schuetzhold+Unruh:WKB-breakdown}).

%%%%%%%%%%%%%%%%%%%%%%%%%%%%%%%%%%%%%%%%%%%%%%%%%%%%%%%%%%%%%%%%%%%%%%%%%%%%%%%%
%%%%%%%%%%%%%%%%%%%%%%%%%%%%%%%%%%%%%%%%%%%%%%%%%%%%%%%%%%%%%%%%%%%%%%%%%%%%%%%%
\section{\label{sec:connection-of-JWKB-solutions}Connection of JWKB solutions
across the event horizon}
%%%%%%%%%%%%%%%%%%%%%%%%%%%%%%%%%%%%%%%%%%%%%%%%%%%%%%%%%%%%%%%%%%%%%%%%%%%%%%%%
%%%%%%%%%%%%%%%%%%%%%%%%%%%%%%%%%%%%%%%%%%%%%%%%%%%%%%%%%%%%%%%%%%%%%%%%%%%%%%%%

In order to continue a mode function $\psi_{\omega}$, which decays
exponentially inside the event horizon (according to the boundary 
condition mentioned above), into the exterior region of the black hole, 
we have to take the full decoupled mode equation~\eqref{eq:decoupled-mode-eq} 
into account. 
The complexity of this equation depends on the concrete pulse shape 
given by $\Omega(\chi)$ and $g(\chi)$. 
As already mentioned in Sec.~\ref{The Model}, we assume that 
$g(\chi)=g_0$ is constant (see also Fig.~\ref{fig:pulse-profile}).
We thus need to specify $\Omega(\chi)$.

We want to model the Schwarzschild metric with 
\begin{equation}
\metric_{00}=1-\frac{2\GN M}{r}\text{ .}
\end{equation}
This metric has a horizon at $r=2\GN M$ (the Schwarzschild radius), 
a singularity at $r=0$, and becomes asymptotically flat for $r\to\infty$. 
For the dielectric black hole analog, the $\metric_{00}$ component behaves as
\begin{equation}
\label{g_00}
\effmetric_{00}\propto 1-\frac{v^2}{\clow^2(\chi)}
=
1-v^2\left(1+\frac{g_0^2}{\Omega^2(\chi)}\right)\text{;}
\end{equation}
see also Eq.~\eqref{eq:effective-metric} below.
Note that $\clow^2(\chi)$ refers to the low-ener\-gy limit 
in Eq.~\eqref{eq:c-low-energy} where the analogy to gravity applies.
In order to model the Schwarzschild metric, we assume the 
following profile (with some constant $\xi>0$):
\begin{equation}
\Omega(\chi)=\Omega_{0}\sqrt{1+2\xi\chi}
\text{ ,}\quad 
g(\chi)=g_0
\text{ .}\label{eq:connection-pulse-profile}
\end{equation}
For simplicity, we choose $\chi$ such that the horizon is located 
at $\chi=0$, so the black hole exterior region is $\chi>0$ and the 
interior region is $\chi<0$.
At the horizon $\chi=0$, we have $\effmetric_{00}=0$, which translates to the
condition~\eqref{eq:horizon-condition}, leading to the relation
\begin{equation}
\Omega_{0}=v\gamma g_{0}\text{ .}\label{eq:Omega-zero}
\end{equation}
The slope $\D\clow(\chi)/\D\chi$ at the horizon determines the 
surface gravity (remember that $v$ is constant), which in turn 
sets the Hawking temperature; see Eqs.~\eqref{eq:Hawking-temp-pls}
and~\eqref{eq:surface-gravity} below.
Similarly to the Schwarzschild metric, where the strength of the 
gravitational field vanishes at $r\to\infty$, the polarizability
of the medium goes to zero as $\chi\to\infty$.
Furthermore, the refractive index diverges (formally) for 
$\chi=-1/(2\xi)$, which is similar to the singularity at $r=0$.
Of course, such a profile~\eqref{eq:connection-pulse-profile} is not a
realistic model for a real laser pulse, but---as we shall see below---it
allows us to derive an exact solution of the mode equation 
in analogy to Ref.~\cite{Schuetzhold+Unruh:Particle-origin}.
Furthermore, because only the vicinity of the horizon is relevant 
for the creation of low-ener\-gy Hawking radiation, we can 
interpret this profile~\eqref{eq:connection-pulse-profile}
as an approximation of a realistic pulse profile in the region 
near the horizon $\xi|\chi|\ll1$.
In terms of a Taylor expansion 
$\Omega^{2}(\chi)=\Omega_{0}^{2}+2\Omega_{0}^{2}\xi\chi+\order(\chi^{2})$,
we neglect the higher-order terms $\order(\chi^{2})$ since $\Omega^{2}(\chi)$
is supposed to be slowly varying.
For example, since the modes we are interested in decay rapidly inside 
the black hole, they do not see the singularity at $\chi=-1/(2\xi)$, 
so that this approximation should be reasonable.
Since the Hawking temperature is supposed to be sufficiently low, 
the JWKB approximation will be valid at both edges 
(inside and outside the black hole) of the linearized region. 
This way, we can continue a JWKB solution inside the black hole across 
the horizon using the linearized mode equation.

Inserting the pulse profile~\eqref{eq:connection-pulse-profile} into the general
decoupled mode equation~\eqref{eq:decoupled-mode-eq} yields the equation
\begin{multline}
\left\{ 
(\omega^{2}+\partial_{\chi}^{2})
\left[(\I\omega+v\partial_{\chi})^{2}+v^{2}g_{0}^{2}\,(1+2\xi\chi)\right]
\vphantom{-g_{0}^{2}\,(\I\omega+v\partial_{\chi})^{2}}
\right.\\
\left.\vphantom{(\omega^{2}+\partial_{\chi}^{2})
\left[(\I\omega+v\partial_{\chi})^{2}+v^{2}g_{0}^{2}\,(1+2\xi\chi)\right]}
-g_{0}^{2}\,(\I\omega+v\partial_{\chi})^{2}\right\} \psi_{\omega}=0
\text{ .}\label{eq:linearized-mode-eq}
\end{multline}
In this section, we will study the solutions of this mode equation in the 
positive frequency range $0<\omega\lesssim\xi$, which covers the essential 
part of the Hawking spectrum. 
The corresponding neg\-a\-tive-fre\-quen\-cy solutions can be derived 
by complex conjugation.
The solution scheme via transformation to momentum space and contour
integration is analogous to Refs.~\cite{Corley:Spectrum-calculation,
Yoshiaki+Takahiro,Saida+Sakagami,Unruh+Schuetzhold:Universality,
Coutant+al:S-matrix-approach,Belgiorno:Hopfield-model-results}, for example.

%%%%%%%%%%%%%%%%%%%%%%%%%%%%%%%%%%%%%%%%%%%%%%%%%%%%%%%%%%%%%%%%%%%%%%%%%%%%%%%%
\subsection{Solution of the mode equation}
%%%%%%%%%%%%%%%%%%%%%%%%%%%%%%%%%%%%%%%%%%%%%%%%%%%%%%%%%%%%%%%%%%%%%%%%%%%%%%%%

Equation~\eqref{eq:linearized-mode-eq} can be solved in reciprocal 
(momentum) space.
To this end, we Fourier--Laplace transform the equation by inserting 
\begin{equation}
\psi_{\omega}(\chi)
=
\intop_{\IC}\tilde{\psi}_{\omega}(k)\E^{\I k\chi}\,\D k
\label{eq:Fourier-Laplace-transform}
\end{equation}
with the (yet unspecified) complex integration contour $\IC$.
By means of this transform, the fourth-order differential 
equation~\eqref{eq:linearized-mode-eq} in $\chi$ becomes the first-order 
differential equation in $k$
($\partial_{\chi}\to\I k$ and $\chi\to\I\partial_{k}$),
\begin{multline}
\!\partial_{k}\tilde{\psi}_{\omega}(k)
=
\frac{1}{2\I v^{2}g_{0}^{2}\xi}
\left[{\left(1+\frac{g_{0}^{2}}{k^{2}-\omega^{2}}\right)}(\omega+vk)^{2}
-v^{2}g_{0}^{2}\right]\\
{}\times\tilde{\psi}_{\omega}(k)\text{ ,}
\label{eq:lin-mode-eq-k-space}
\end{multline}
which is easy to solve for $\tilde{\psi}_{\omega}(k)$.
After transforming back to position space via 
Eq.~\eqref{eq:Fourier-Laplace-transform}, the solution reads
\begin{equation}
\psi_{\omega}(\chi)
=
\intop_{\IC}f(k)\E^{\chi h(k)}\,\D k\text{ ,}
\label{eq:mode-func-contour-integral}
\end{equation}
with the two auxiliary functions
\begin{align}
f(k)={}&
\left[\I\left(\frac{k}{\omega}-1\right)\right]^{-\I\,(1+v)^2 \omega/(4v^2 \xi)}
\nonumber \\
&{}\times
\left[\I\left(\frac{k}{\omega}+1\right)\right]^{\I\,(1-v)^2 \omega/(4v^2 \xi)}
\label{eq:cont-int-aux-func-one}
\end{align}
and
\begin{equation}
h(k)=\I k
\left(1-\frac{v^{2}k^{2}/3+v\omega k+\omega^{2}}{2v^{2}g_{0}^{2}\xi\chi}\right)
\text{.}\label{eq:cont-int-aux-func-two}
\end{equation}
The constant of integration, which is irrelevant for the Hawking spectrum,
is related to the contour $\IC$ and has been omitted for simplicity.
Note that the exponent in Eq.~\eqref{eq:mode-func-contour-integral} can be cast
into the form
\begin{equation}
\chi h(k)=\I k\chi-\I\,\frac{(k+\omega/v)^3}{6g_0^2\xi}
+\I\,\frac{\omega^3}{6 v^3 g_0^2 \xi}\text{ ,}
\end{equation}
where the last term can be absorbed by the integration constant
mentioned above.

Equation~\eqref{eq:mode-func-contour-integral} is a contour integral 
representation of the mode function $\psi_{\omega}$. 
In order to define complex powers as in $f(k)$, we have to specify the 
two branch cuts of the complex natural logarithm starting at the two 
singularities $k=\pm\omega$.
Here, we choose these branch cuts to run upwards in the complex plane 
(parallel with respect to the positive imaginary axis).
This corresponds to the principal value $\Ln\mathfrak{z}$ of the complex 
natural logarithm (i.e., $-\pi<\Im[\Ln\mathfrak{z}]\leq\pi$).  
As we shall see below, this choice is most convenient for deriving 
mode functions which satisfy the required boundary condition, 
that is, are evanescent inside the black hole.
With other choices, we can derive other solutions of the wave equation
(e.g., with a contribution from the partner mode inside).

\boldmath
%%%%%%%%%%%%%%%%%%%%%%%%%%%%%%%%%%%%%%%%%%%%%%%%%%%%%%%%%%%%%%%%%%%%%%%%%%%%%%%%
\subsection{Mode function inside the black hole\texorpdfstring{ ($\chi<0$)}{}:
Boundary condition}
%%%%%%%%%%%%%%%%%%%%%%%%%%%%%%%%%%%%%%%%%%%%%%%%%%%%%%%%%%%%%%%%%%%%%%%%%%%%%%%%
\unboldmath 

Let us first consider the mode function $\psi_{\omega}$ in 
Eq.~\eqref{eq:mode-func-contour-integral} inside the black hole, where
we have imposed the boundary condition for the derivation of the 
Hawking effect according to Sec.~\ref{sub:JWKB-modes-inside}. 
This boundary condition is fulfilled by selecting the 
(end points of the) integration contour $\IC$ appropriately. 
The contour can ``safely'' run to infinity into any direction of the 
complex plane where the exponential part of the integrand, 
$\exp[\chi h(k)]$, decays to zero; that is,
$\Re[\chi h(k)]\to-\infty$, while the function $f(k)$ is unproblematic.
These ``valleys'' of the integrand are located at 
\begin{itemize}
 \item $\pi/3<\Arg k<2\pi/3$ (top), 
 \item $-\pi<\Arg k<-2\pi/3$ (bottom left), and 
 \item $-\pi/3<\Arg k<0$ (bottom right). 
\end{itemize}
For other directions (i.e., between these valleys),  
$|\exp[\chi h(k)]|$ diverges for $|k|\to\infty$.

\begin{figure}
\subfloat[\label{fig:Cin}Integration contour $\ICin$ (schematic).]%
{\includegraphics{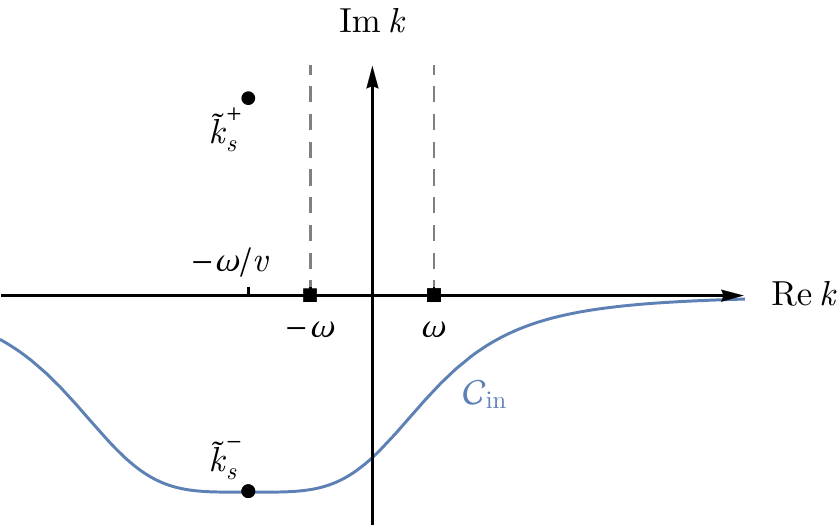}}

\subfloat[\label{fig:integrand-inside}Absolute value (logarithmic) of the
integrand.]{\includegraphics{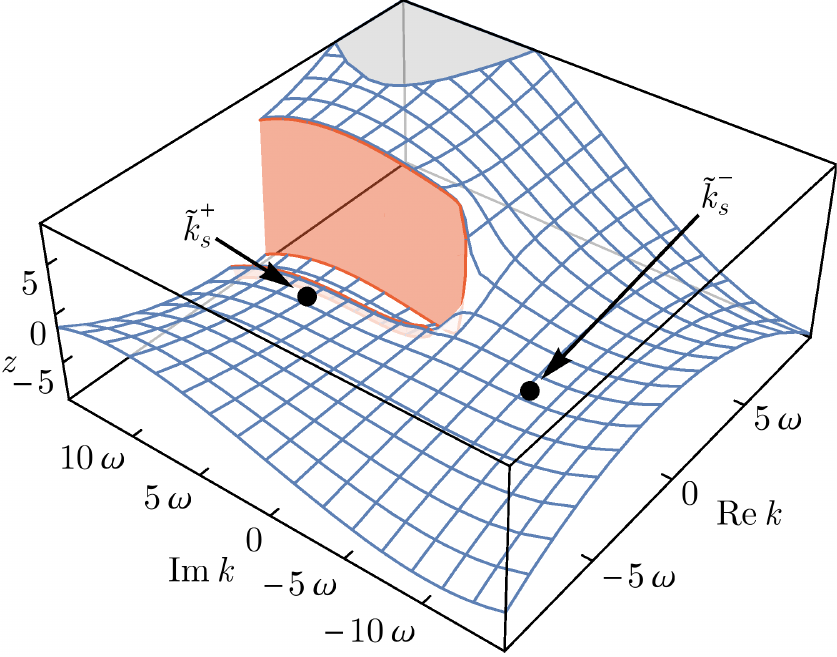}}

\caption{\label{fig:contour-integration-inside}%
Contour integration~\protect\eqref{eq:mode-func-contour-integral}
inside the black hole ($\chi<0$).
(\protect\subref*{fig:Cin})~Integration contour $\ICin$ running through 
the saddle point $\ktildesminus$ (not drawn to scale).
The squares mark the singularities of the integrand due to the 
complex powers in $f(k)$, and the dashed lines are the 
corresponding branch cuts.
(\protect\subref*{fig:integrand-inside})~%
Logarithmic landscape plot of the absolute value of the integrand,
i.e., $z=\ln|f(k)\E^{\chi h(k)}|$ over a complex $k$. 
The parameter values applied in this plot ($v=1/2$, 
$\xi=\omega=g_{0}/10$, and $\chi=-3/g_{0}$, so $\Omega\approx0.4g_{0}$) 
are not within the validity range~\eqref{eq:SP-method-validity-condition} of
the saddle point approximation but have been chosen for illustrative purposes.
We see that the saddle point $\ktildesminus$ connects the 
two valleys of the integrand in the lower complex half plane.}
\end{figure}

Here, we choose to integrate just below the real axis from $-\infty$ to
$\infty$, that is, from the bottom-left valley into the bottom-right
one and below the singularities at $k=\pm\omega$, so we have fixed
the contour $\IC$ appearing in Eqs.~\eqref{eq:Fourier-Laplace-transform}
and~\eqref{eq:mode-func-contour-integral}.
We can still, however, deform the integration contour by means of 
Cauchy's theorem in order to simplify the integration without changing 
the value of the integral.
The exponent function $h(k)$ in the integrand has two saddle points,
which satisfy $h^\prime (\ks)=0$.
For $\chi<0$, these saddle points are 
\begin{equation}
\ktildesplusminus=\pm\I g_{0}\sqrt{-2\xi\chi}-\frac{\omega}{v}
\text{ .}\label{eq:saddle-points-inside}
\end{equation}
The saddle point $\ktildesminus$ connects the two adjacent valleys in the lower
complex half plane (bottom left and right) with each other.
Hence, we may deform $\IC$ smoothly (keeping the end points fixed) 
without ever encountering any singularities or branch cuts of the
integrand so that the final contour $\ICin$ runs through $\ktildesminus$
along the ``mountain pass route''; see
Fig.~\ref{fig:contour-integration-inside}.
The contribution from the saddle point $\ktildesminus$ will dominate the value 
of the integral since the rest of the integration runs through the valleys.

The other saddle point $\ktildesplus$ corresponds to a ``mountain pass''
which connects the upper (top) valley with the two lower (bottom) valleys 
with a bifurcation point near the origin.
The height of this pass increases exponentially for increasing $|\chi|$, but our
selected contour does not go through this point.

%%%%%%%%%%%%%%%%%%%%%%%%%%%%%%%%%%%%%%%%%%%%%%%%%%%%%%%%%%%%%%%%%%%%%%%%%%%%%%%%
%%%%%%%%%%%%%%%%%%%%%%%%%%%%%%%%%%%%%%%%%%%%%%%%%%%%%%%%%%%%%%%%%%%%%%%%%%%%%%%%

Applying the saddle point method (see, e.g., Ref.~\cite{Wong:Asymptotics}), the
mode function 
$\psi_{\omega}(\chi<0)\approx\psi_{\omega}^{\mathrm{inside}}(\chi)$
inside the black hole is thus approximately given by the saddle point 
contribution from $\ktildesminus$.
The lead\-ing-ord\-er term of the saddle point expansion 
(which becomes asymptotically exact in the formal limit $\chi\to-\infty$) 
reads
\begin{equation}
\psi_{\omega}^{\mathrm{inside}}(\chi)
=
\sqrt{\frac{2\pi}{|\chi h''(\ktildesminus)|}}
f(\ktildesminus)\E^{\chi h(\ktildesminus)}
\text{ .}\label{eq:SP-contrib-inside}
\end{equation}
The higher-order corrections of the saddle point expansion are negligible if
\begin{equation}
|\xi\chi|^{3/2}\gg\frac{\gamma}{v}\frac{\xi}{\Omega_{0}}
=
\frac{\xi}{v^{2}g_{0}}
\text{ .}\label{eq:SP-method-validity-condition}
\end{equation}
For the derivation of this inequality, we have used that the typical frequency
of Hawking radiation $\omega\lesssim\order(\xi)$ is set by the surface gravity.
Since $g_0$ and $\Omega_0$ are characteristic scales of the 
medium and thus are supposed to be much larger than $\xi$ 
(and $\omega$), the right-hand side of 
Eq.~\eqref{eq:SP-method-validity-condition} is very small.
Sufficiently far inside the black hole as determined by the 
condition~\eqref{eq:SP-method-validity-condition}, 
the mode function $\psi_{\omega}(\chi<0)$ is therefore approximated
well by the lead\-ing-ord\-er term~\eqref{eq:SP-contrib-inside}.
However, we want to stay away from the singularity, so we assume 
$\xi|\chi|\ll1$.
Hence $\xi$ has to be small enough so that both assumptions can 
be satisfied simultaneously (low Hawking temperature).

Let us check to see whether $\psi_{\omega}^{\mathrm{inside}}$ does indeed
satisfy the required boundary condition. 
We evaluate the absolute value of Eq.~\eqref{eq:SP-contrib-inside}
to find
\begin{align}
|\psi_{\omega}^{\mathrm{inside}}(\chi)| 
& =
\sqrt[4]{\frac{2\pi^{2}g_{0}^{2}\xi}{|\chi|}}|f(\ktildesminus)|
\E^{-2g_{0}|\chi|\sqrt{2\xi|\chi|}/3}
\nonumber \\
& \propto
\frac{\E^{-2g_{0}|\chi|\sqrt{2\xi|\chi|}/3}}{\sqrt[4]{|\chi|}}
\text{ ,}
\end{align}
so the mode function decays exponentially beyond the horizon, and
hence the integration contour is in accordance with the boundary condition.
For values of $\chi$ satisfying the 
condition~\eqref{eq:SP-method-validity-condition}, we thus find that 
$|\psi_{\omega}^{\mathrm{inside}}(\chi)|$ is suppressed exponentially. 

Note that the contribution from the other saddle point $\ktildesplus$
would instead grow exponentially when going further and further inside 
the black hole---which explains why we selected our contour in such a 
way that it does not go through this saddle point.

\boldmath
%%%%%%%%%%%%%%%%%%%%%%%%%%%%%%%%%%%%%%%%%%%%%%%%%%%%%%%%%%%%%%%%%%%%%%%%%%%%%%%%
\subsection{Mode function outside the black hole\texorpdfstring{ ($\chi>0$)}{}:
Identification of the JWKB modes}
%%%%%%%%%%%%%%%%%%%%%%%%%%%%%%%%%%%%%%%%%%%%%%%%%%%%%%%%%%%%%%%%%%%%%%%%%%%%%%%%
\unboldmath

\begin{figure}
\subfloat[\label{fig:Cout}Integration contour $\ICout$ (schematic).]%
{\includegraphics{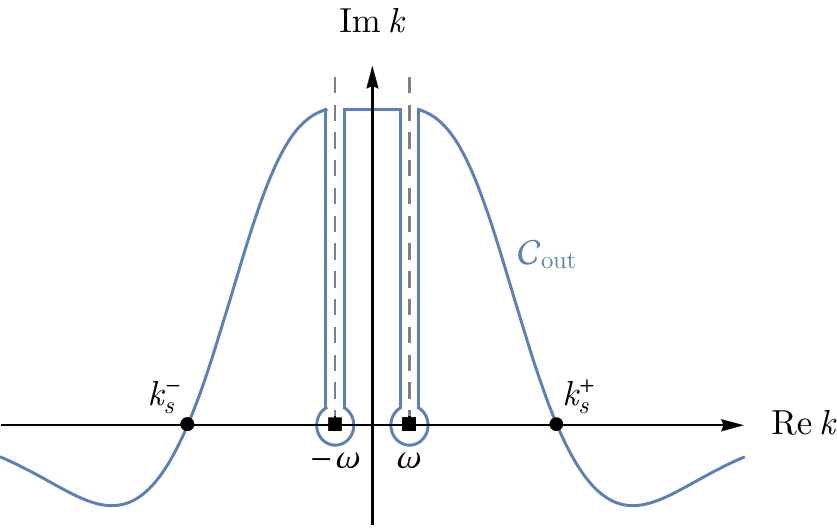}}

\subfloat[\label{fig:integrand-outside}Absolute value (logarithmic) of the
integrand.]{\includegraphics{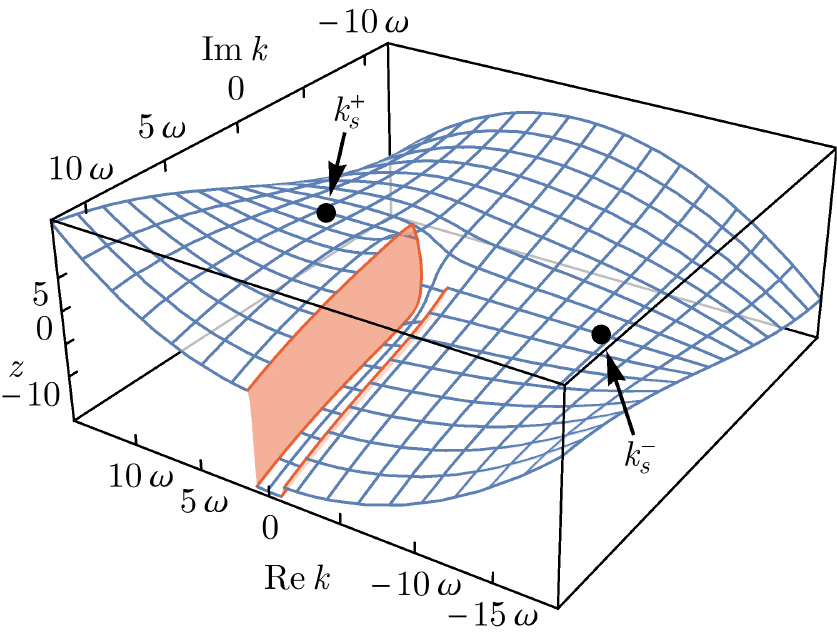}}

\caption{\label{fig:contour-integration-outside}%
Contour integration~\protect\eqref{eq:mode-func-contour-integral}
outside the black hole ($\chi>0$).
(\protect\subref*{fig:Cout})~%
Deformed integration contour $\ICout$ running through the saddle points 
$\ksplusminus$ and circumventing the branch cuts. 
This sketch is not drawn to scale and the saddle points are drawn symmetrically
around zero (not $-\omega/v$) for simplicity.
(\protect\subref*{fig:integrand-outside})~%
Landscape plot of the integrand, i.e., $z=\ln|f(k)\E^{\chi h(k)}|$ over 
$k\in\mathbb{C}$, for the same system as in
Fig.~\protect\subref{fig:integrand-inside} ($v=1/2$, $\xi=\omega=g_{0}/10$) but
outside the horizon at $\chi=5/g_{0}$, where $\Omega\approx0.8g_{0}$.
Note that the discontinuity of the branch cut at $\Re k=+\omega$ is
significantly larger than that at $\Re k=-\omega$, which makes the latter a bit
hard to notice.}
\end{figure}

Now that we know the correct integration contour, we evaluate the contour 
integral~\eqref{eq:mode-func-contour-integral} outside the black hole in 
order to calculate the analytically continued mode function 
$\psi_{\omega}(\chi>0)$.
However, for $\chi>0$, the saddle points of $h(k)$ have moved to the positions
\begin{equation}
\ksplusminus=\pm g_{0}\sqrt{2\xi\chi}-\frac{\omega}{v}
\label{eq:saddle-points-outside}
\end{equation}
on the real axis, so $\ICin$ is not the most advantageous integration contour
for $\chi>0$. 
The validity condition~\eqref{eq:SP-method-validity-condition}
for the saddle point approximation (which we want to apply again)
implies that $\chi$ must be large enough that 
$|\ksplusminus|\gg\omega$, which means that the singularities of the integrand
at $k=\pm\omega$ are between $\ksminus$ and $\ksplus$. 
The saddle point $\ksminus$ connects the bottom-left valley (where the
integration contour starts) with the top valley in the upper complex half plane.
The other saddle point $\ksplus$ leads from this valley into the bottom-right 
valley where the integration ends. Hence, we deform $\ICin$ again, always
avoiding going across any non\-holo\-morphic regions of the integrand,
so that the final contour $\ICout$ follows the path of steepest descent
through the saddle points.
The top valley, which $\ICout$ must traverse, however, is divided by the branch
cuts, so the contour must circumvent these two discontinuous half lines in the
complex plane as depicted in Fig.~\ref{fig:contour-integration-outside}.
Note that, in contrast to Ref.~\cite{Belgiorno:Hopfield-model-results},
we do not neglect any branch cuts.
Putting all dominant contributions to the integral together, the mode function
outside the black hole is thus composed of the saddle point contributions
$\psi_{\omega}^{\pm}$ and the functions $\psicutpm_{\omega}$, which
are due to the circumvention of the branch cuts originating from $k=\pm\omega$,
\begin{equation}
\psi_{\omega}(\chi>0)
\approx
\psi_{\omega}^{-}(\chi)+\psicutm_{\omega}(\chi)+\psicutp_{\omega}(\chi)
+\psi_{\omega}^{+}(\chi)
\text{ .}
\end{equation}
Let us now identify the four JWKB modes which have been explained
in Sec.~\ref{sub:JWKB-modes-outside} in this mode function.

%%%%%%%%%%%%%%%%%%%%%%%%%%%%%%%%%%%%%%%%%%%%%%%%%%%%%%%%%%%%%%%%%%%%%%%%%%%%%%%%
%%%%%%%%%%%%%%%%%%%%%%%%%%%%%%%%%%%%%%%%%%%%%%%%%%%%%%%%%%%%%%%%%%%%%%%%%%%%%%%%

The lead\-ing-order saddle point contributions read 
\begin{multline}
\psi_{\omega}^{\pm}(\chi)
=
\E^{\mp\I\pi/4}\,\sqrt[4]{\frac{2\pi^{2}g_{0}^{2}\xi}{\chi}}\,
\E^{\chi h(\ksplusminus)}\,
\E^{\pm\pi\omega/(2v\xi)}
\\
\times\left|\frac{\ksplusminus}{\omega}-1\right|^{-\I(1+v)^2 \omega/(4v^2 \xi)}
\left|\frac{\ksplusminus}{\omega}+1\right|^{\I(1-v)^{2}\omega/(4v^{2}\xi)}
\text{.}\label{eq:SP-contrib-outside}
\end{multline}
These functions oscillate rapidly due to the large saddle point 
values $|\ksplusminus|\sim g_{0}\sqrt{2\xi\chi}\gg\omega$
as implied by inequality~\eqref{eq:SP-method-validity-condition}
and $\xi\chi\ll1$.
For large $\chi$, these expressions become exact solutions of the 
decoupled mode equation~\eqref{eq:linearized-mode-eq}, 
so they are asymptotic independent modes and should
therefore coincide with two JWKB modes in this limit.
Solving the local dispersion relation~\eqref{eq:disp-rel-polynomial} 
for the current pulse profile~\eqref{eq:connection-pulse-profile} 
and large wave vectors (neglecting $\omega$) yields two solutions 
$\JWKBk\approx\pm g_{0}\sqrt{2\xi\chi}$, which are asymptotically
equal to the saddle points $\ksplusminus$.
Hence, the modes $\psi_{\omega}^{\pm}$ correspond (asymptotically)
to the JWKB modes $\kstarplusminus$.

Now we consider the branch cut contributions.
The contour $\ICout$ may run arbitrarily deep into the top valley 
(where the integrand is exponentially suppressed), so we integrate 
infinitesimally close on both sides along the cuts, respectively, 
up to infinity.
The small circles around the singularities in Fig.~\subref{fig:Cout} do not
yield any contributions.
After some substitutions, the resulting (exact) integrals can be written
\begin{multline}
\psicutpm_{\omega}(\chi)
=
\pm2\I\sinh\!\left[\frac{\pi\,(1\pm v)^{2}\omega}{4v^{2}\xi}\right]
\intop_{0}^{\infty}
\left(\frac{u}{\omega}\right)^{\mp\I\frac{(1\pm v)^{2}\omega}{4v^{2}\xi}}
\\
\times\left(-\frac{u}{\omega}\pm2\I\right)^{\pm\I(1\mp v)^{2}\omega/(4v^{2}\xi)}
\E^{\chi h(\I u\pm\omega)}\,\D u\text{ .}
\label{eq:cut-contrib-exact}
\end{multline}
As with the saddle point contributions, we can uniquely identify 
$\psicutpm_{\omega}$ with JWKB modes, respectively, in the limit 
$\chi\to\infty$.
The integrand in Eq.~\eqref{eq:cut-contrib-exact} can be substantially
simplified for a large $\chi$ because then only small values $u\ll\omega$
are significant for the integration.
We can express the remaining integral by the gamma function $\GammaFunc$.
The resulting mode functions read
\begin{multline}
\psicutpm_{\omega}(\chi)
\sim
\omega\,(\pm2\I)^{1\pm\I(1\mp v)^{2}\omega/(4v^{2}\xi)}
\sinh\!\left[\frac{\pi\,(1\pm v)^{2}\omega}{4v^{2}\xi}\right]
\\
\times\GammaFunc\!\left[1\mp\frac{\I\,(1\pm v)^{2}\omega}{4v^{2}\xi}\right]
\left(\frac{1}{\omega\chi}\right)^{1\mp\I(1\pm v)^{2}\omega/(4v^{2}\xi)}
\E^{\pm\I\omega\chi}
\text{ .}\label{eq:cut-contrib-asymptotic}
\end{multline}
This is the lead\-ing-order asymptotic term of the exact integral
in Eq.~\eqref{eq:cut-contrib-exact}, so it consequently solves the
decoupled mode equation~\eqref{eq:linearized-mode-eq} in the limit
$\chi\to\infty$.
These functions do still solve the mode equation for $\chi\to\infty$ 
if dispersion is neglected, that is, if we discard all terms containing 
third- or higher-order derivatives acting on the $\psi$ field 
($\omega^{n}\partial_{\chi}^{m}$, with $n+m>2$ since
$\omega$ originates from a time derivative $\partial_{\tau}$).
The modes described by $\psicutpm_{\omega}$ are thus only slightly affected
by dispersion far outside the black hole and correspond to the Hawking
mode $\kstarH$ and the counterpropagating mode $\kstarcp.$
The phase/group velocity of $\psicutp_{\omega}$ is positive, so this mode 
propagates away from the black hole and can therefore be identified with 
the Hawking mode.
The other branch cut contribution $\psicutm_{\omega}$ has a negative 
group velocity, so it corresponds to the counterpropagating mode.

%%%%%%%%%%%%%%%%%%%%%%%%%%%%%%%%%%%%%%%%%%%%%%%%%%%%%%%%%%%%%%%%%%%%%%%%%%%%%%%%
\subsection{Current densities}
%%%%%%%%%%%%%%%%%%%%%%%%%%%%%%%%%%%%%%%%%%%%%%%%%%%%%%%%%%%%%%%%%%%%%%%%%%%%%%%%

For the derivation of the Hawking spectrum, we need to know the contribution
from the pos\-i\-tive- and neg\-a\-tive-norm in-modes ($\kstarplus$ and
$\kstarminus$ in the JWKB picture) to the outgoing Hawking radiation.
In the time-in\-de\-pend\-ent limit (stationary modes), we therefore have 
to calculate the Noether charge current densities $j_{\omega}^{\pm}$ of the 
modes $\psi_{\omega}^{\pm}$ using Eqs.~\eqref{eq:mode-func-A-from-psi} and~%
\eqref{eq:Noether-current-density}, with $\partial_{\tau}\to-\I\omega$.
The current density of a stationary mode is exactly constant 
(see Sec.~\ref{sub:Noether-charge-conservation}),
so we may calculate $j_{\omega}^{\pm}$ very far outside the black hole
($\chi\to\infty$) where all $\chi$-depend\-ent terms in $j_{\omega}^{\pm}$
vanish.
(These terms are artifacts caused by the saddle point approximation anyway.)
The resulting current densities assume the simple form
\begin{equation}
j_{\omega}^{\pm}
=
\mp2\pi\Omega_{0}^{2}\xi\,\E^{\pm\pi\omega/(v\xi)}
\text{ .}\label{eq:j-plusminus}
\end{equation}
The current density $\jcutp_{\omega}$ associated with the branch cut
contribution $\psicutp_{\omega}$ (Hawking mode), which we also need to know for
deriving the Hawking spectrum, is calculated in the same way as
$j_{\omega}^{\pm}$ above.
The result reads
\begin{equation}
\jcutp_{\omega}
=
2\pi\Omega_{0}^{2}\xi\left[
\E^{\pi\omega/(v\xi)}-\E^{-\pi\,(1+v^{2})\omega/(2v^{2}\xi)}
\right]\text{.}\label{eq:j-cut-plus}
\end{equation}
Of course, the current density of the exponentially decaying mode 
inside the black hole vanishes.

%%%%%%%%%%%%%%%%%%%%%%%%%%%%%%%%%%%%%%%%%%%%%%%%%%%%%%%%%%%%%%%%%%%%%%%%%%%%%%%%
%%%%%%%%%%%%%%%%%%%%%%%%%%%%%%%%%%%%%%%%%%%%%%%%%%%%%%%%%%%%%%%%%%%%%%%%%%%%%%%%
\section{Hawking spectrum}
%%%%%%%%%%%%%%%%%%%%%%%%%%%%%%%%%%%%%%%%%%%%%%%%%%%%%%%%%%%%%%%%%%%%%%%%%%%%%%%%
%%%%%%%%%%%%%%%%%%%%%%%%%%%%%%%%%%%%%%%%%%%%%%%%%%%%%%%%%%%%%%%%%%%%%%%%%%%%%%%%

We are now in the position to calculate the Hawking spectrum.
Here, we only present a brief review of the derivation of the Hawking effect.
For a more detailed explanation, we refer the reader to
Ref.~\cite{Corley:Sonic-dispersion}, for example.

The Hawking effect requires a quan\-tum-field-theo\-ret\-ic framework, so the
classical fields $A$ and $\psi$ are substituted by Hermitian field operators
$\hat{\mathfrak{A}}$ and $\hat{\Psi}$, which obey the bosonic equal-time
commutation relations and solve the same equations of motion as the classical
fields.
Hence, the Klein--Gordon inner product~\eqref{eq:inner-product} continues to be
useful in the context of the quantized fields.
We find
\begin{multline}
\left[
\IP{\begin{pmatrix}A_1\\ \psi_1\end{pmatrix}}
{\begin{pmatrix}\hat{\mathfrak{A}}\\ \hat{\Psi}\end{pmatrix}},
\IP{\begin{pmatrix}A_2\\ \psi_2\end{pmatrix}}
{\begin{pmatrix}\hat{\mathfrak{A}}\\ \hat{\Psi}\end{pmatrix}}
\right]\\
=-\IP{\begin{pmatrix}A_1\\ \psi_1\end{pmatrix}}
{\begin{pmatrix}A_2^{\ast}\\ \psi_2^{\ast}\end{pmatrix}}
\label{eq:inner-product-commutator}
\end{multline}
(cf.\ Ref.~\cite{Corley:Sonic-dispersion}), where $[\cdot,\cdot]$ denotes
the commutator and $(A_n,\psi_n)$ are two arbitrary, classical
field solutions of the equations of motion. 
The inner product allows us to ``project'' the field operators onto a set 
of classical field solutions (mode expansion), and 
Eq.~\eqref{eq:inner-product-commutator}
facilitates the derivation of the corresponding annihilation and creation
operators $\aOp$ and $\aOp^{\dagger}$.
As one may infer from the above commutator, pos\-i\-tive-norm modes 
correspond to annihilation operators, while neg\-a\-tive-norm modes 
correspond to creation operators.

As explained in Sec.~\ref{sub:JWKB-modes-inside}, a late-time outgoing
Hawking wave (packet) originates from contributions from the three
in-modes $\kstarplusminus$ and $\kstarcp$ at early times.
For stationary modes (time-inde\-pend\-ent picture) and expressed via 
creation and annihilation operators, this statement reads
\begin{equation}
\aH_{\omega}
=
\alpha_{\omega}\aplus_{\omega}+\beta_{\omega}(\aminus_{\omega})^{\dagger}
+\eta_{\omega}\acp_{\omega}
\label{eq:Bogoliubov-trafo}
\end{equation}
with the Bogoliubov coefficients $\alpha_{\omega}$, $\beta_{\omega}$,
and $\eta_{\omega}$.
As explained above, the rapidly oscillating, neg\-a\-tive-norm mode
is represented by a creation operator $(\aminus_{\omega})^{\dagger}$ 
in this relation.
Since the above operators obey the usual commutation relations for bosonic
creation and annihilation operators, we obtain the (unitarity) relation
\begin{equation}
\label{unitarity}
|\alpha_{\omega}|^2-|\beta_{\omega}|^2+|\eta_{\omega}|^2=1
\text{ .}
\end{equation}
Note that this equality can also be derived from the properties 
of the classical solutions of the wave equation. 
In terms of the current densities, this relation corresponds to 
charge conservation 
\begin{equation}
j_{\omega}^{+}+j_{\omega}^{-}+\jcutm_{\omega}+\jcutp_{\omega}=0
\text{ .}
\end{equation}
Here the saddle point contribution $j_{\omega}^{+}$ from $\ksplus$ 
corresponds to $|\alpha_{\omega}|^2$, while the other one, $j_{\omega}^{-}$
from $\ksminus$, is associated with $|\beta_{\omega}|^2$.
Furthermore, the two branch cut contributions $\jcutm_{\omega}$ 
and $\jcutp_{\omega}$ correspond to the counterpropagating mode 
(i.e., $|\eta_{\omega}|^2$) and the Hawking mode, respectively.
Note that $\rho_\omega^-$ is negative, while the other three, 
$\rho_\omega^+$, $\rho_{\omega}^{\mathrm{cut}{-}}$, and
$\rho_{\omega}^{\mathrm{cut}{+}}$, are positive; see
Eq.~\eqref{Noether-charge-sign}.
However, since only the Hawking mode $\psicutp_{\omega}$
has a positive group velocity (away from the horizon) 
while the other three are negative (towards the horizon),
we find that $\jcutp_{\omega}$ and $j_{\omega}^{-}$ are positive, while 
$\jcutm_{\omega}$ and $j_{\omega}^{+}$ are negative.
Altogether, with the correct normalization, we have the following 
identifications:
\begin{itemize}
 \item $|\alpha_{\omega}|^2\to-j_{\omega}^{+}/\jcutp_{\omega}$,
 \item $|\beta_{\omega}|^2\to+j_{\omega}^{-}/\jcutp_{\omega}$, and
 \item $|\eta_{\omega}|^2\to-\jcutm_{\omega}/\jcutp_{\omega}$.
\end{itemize}
We assume the in-vacuum state for the fields in the dielectric medium.
This quantum state is defined by
\begin{equation}
\aplus_{\omega}\ket{0_\initial}=\aminus_{\omega}\ket{0_\initial}
=\acp_{\omega}\ket{0_\initial}=0
\text{ ,}
\end{equation}
so there are no particles initially.
Using Eq.~\eqref{eq:Bogoliubov-trafo}, the mean number of Hawking particles 
emitted (per unit time) from the in-vacuum state turns out to be
\begin{equation}
\Braket{0_\initial|(\aH_\omega)^{\dagger}\aH_\omega|0_\initial}
=
\Braket{\nH_\omega}_\initial
=
|\beta_{\omega}|^{2}
\text{ ,}\label{eq:Hawking-spectrum-Bogoliubov-coeff}
\end{equation}
which is the quantity we are interested in.

For stationary modes, Eq.~\eqref{eq:Hawking-spectrum-Bogoliubov-coeff}
can be evaluated by means of the current densities $j_{\omega}$ of
the individual modes, which describe the propagation of the conserved
Noether charge.
The Hawking particle yield is given by the relative amount of negative 
charge contribution $j_{\omega}^{-}$ from the neg\-a\-tive-norm in-mode 
to the outgoing Hawking flux $\jcutp_{\omega}$.
We already calculated these current densities; see Eqs.~\eqref{eq:j-plusminus}
and~\eqref{eq:j-cut-plus}.
From Eq.~\eqref{eq:j-plusminus} and the above identification, we may infer
\begin{equation}
\label{ratio}
\left|\frac{j_{\omega}^{-}}{j_{\omega}^{+}}\right|
=
\frac{|\beta_{\omega}|^2}{|\alpha_{\omega}|^2}
=
\exp\!\left(-\frac{2\pi\omega}{v\xi}\right)\text{.}
\end{equation}
Exploiting the unitarity relation~\eqref{unitarity}, we find 
\begin{equation}
|\beta_{\omega}|^{2}
=
\frac{1-|\eta_{\omega}|^2}{\E^{2\pi\omega/(v\xi)}-1}
= 
\frac{\GBF_{\omega}}{\E^{\omega/\THawking}-1}
\label{eq:Hawking-spectrum-pls}
\end{equation}
with the Hawking temperature
\begin{equation}
\THawking=\frac{v\xi}{2\pi}
\label{eq:Hawking-temp-pls}
\end{equation}
and the fre\-quen\-cy-depend\-ent gray-body factor
\begin{equation}
\GBF_\omega=1-|\eta_{\omega}|^2
\text{ ,}\label{eq:gray-body-factor-pls}
\end{equation}
which can be determined by comparing $j_{\omega}^{-}/\jcutp_{\omega}$ 
(yielding $|\beta_{\omega}|^2$) with the above expression:
\begin{equation}
\GBF_\omega
=
\frac{2\sinh[\pi\omega/(v\xi)]}
{\E^{\pi\omega/(v\xi)}-\E^{-\pi(1+v^2)\omega/(2v^2\xi)}}
\text{ .}
\end{equation}
As expected, this factor is bounded from above and below via 
$0<\GBF_\omega<1$ and approaches unity for $\omega/\xi\to\infty$ 
and also for $v\to1$.
For $\omega/\xi\to0$, it converges to a finite value $4v/(1+v)^2<1$.
For a small $\omega$, the spectrum thus behaves as $1/\omega$---the same
scaling that was found in Ref.~\cite{Finazzi:Hopfield-model-results} for a
step-func\-tion profile.

%%%%%%%%%%%%%%%%%%%%%%%%%%%%%%%%%%%%%%%%%%%%%%%%%%%%%%%%%%%%%%%%%%%%%%%%%%%%%%%%
\subsection{Transformation to the laboratory frame}
%%%%%%%%%%%%%%%%%%%%%%%%%%%%%%%%%%%%%%%%%%%%%%%%%%%%%%%%%%%%%%%%%%%%%%%%%%%%%%%%

As a final step, we derive the frequency spectrum of Hawking radiation
as measured in the laboratory.
We thus have to express the pulse frame quantities $\omega$ and $\xi$ 
in terms of laboratory frame quantities.

Let us start with the frequency $\omega$.
Asymptotically (large $\chi$), a Hawking wave with the frequency $\omega>0$
in the pulse frame oscillates with the wave vector $k=+\omega$;
see for example the functional form of $\psicutp_{\omega}$ in 
Eq.~\eqref{eq:cut-contrib-asymptotic}.
According to the Lorentz boost~\eqref{eq:Lorentz-boost},
this wave has the frequency
\begin{equation}
\omega_{\mathrm{lab}}
=
\gamma\left(1+v\right)\omega
\label{eq:spectrum-trafo-frequencies}
\end{equation}
in the laboratory frame.
This equation allows us to express $\omega$ in the Hawking spectrum 
in terms of $\omega_{\mathrm{lab}}$.

The sur\-face-grav\-i\-ty-like quantity $\xi$ has been defined
as the value of $(\partial_{\chi}\Omega)/\Omega$ at the horizon $\chi=0$
in Sec.~\ref{sec:connection-of-JWKB-solutions}.
Hence, the corresponding quantity $\xi_{\mathrm{lab}}$ in the 
laboratory frame (according to which the horizon is located at $x=vt$)
is Lorentz contracted, as proven by
\begin{equation}
\xi_{\mathrm{lab}}
=
\left.\frac{\partial_{x}\Omega}{\Omega}\right|_{x-vt=0}=\gamma\xi\text{ ,}
\end{equation}
where we have inserted the pulse shape~\eqref{eq:connection-pulse-profile}
in laboratory frame coordinates.

%%%%%%%%%%%%%%%%%%%%%%%%%%%%%%%%%%%%%%%%%%%%%%%%%%%%%%%%%%%%%%%%%%%%%%%%%%%%%%%%
%%%%%%%%%%%%%%%%%%%%%%%%%%%%%%%%%%%%%%%%%%%%%%%%%%%%%%%%%%%%%%%%%%%%%%%%%%%%%%%%
\section{Effective Metric}
%%%%%%%%%%%%%%%%%%%%%%%%%%%%%%%%%%%%%%%%%%%%%%%%%%%%%%%%%%%%%%%%%%%%%%%%%%%%%%%%
%%%%%%%%%%%%%%%%%%%%%%%%%%%%%%%%%%%%%%%%%%%%%%%%%%%%%%%%%%%%%%%%%%%%%%%%%%%%%%%%

Now let us discuss the analogy to gravity.
If we assume slowly varying fields and thus neglect higher-order time
derivatives, we may insert the approximate solution
$\psi\approx g\partial_t A/\Omega^2$ back into the original
action~\eqref{eq:Llab} and obtain the low-ener\-gy effective action for the
macroscopic electromagnetic field (in the laboratory frame):
\begin{equation}
\Leff
=
\frac{1}{2}\left[\left(1+\frac{g^2}{\Omega^2}\right)|\partial_{t}A|^{2}
-|\partial_{x}A|^{2}\right]\text{.}
\label{eq:Leff}
\end{equation}
As expected, the low-ener\-gy effective equation of motion from this action
reproduces the dispersion relation~\eqref{eq:c-low-energy}.
This action is analogous to that of a scalar field $A(t,x)$
in the effective metric
\begin{equation}
\D\proptimeeff^2=\D t^2-
\left[1+\frac{g^2}{\Omega^2}\right]\left(\D x^2+\D y^2\right) 
\text{.}
\end{equation}
Note that an effective metric in 1+1 dimensions would not be sufficient 
(unless $g/\Omega$ is constant) because the effective action~\eqref{eq:Leff} 
is not conformally invariant.
Furthermore, the above form is not unique---one could also use other choices
(e.g., in 3+1 dimensions).

After a Lorentz boost to the pulse frame according to
Eq.~\eqref{eq:Lorentz-boost}, the effective metric transforms to 
\begin{align}
\label{eq:effective-metric}
\D\proptimeeff^2={}&
\gamma^2\left(1-v^2\left[1+\frac{g^2}{\Omega^2}\right]\right)\D\tau^2
-2v\gamma^2\frac{g^2}{\Omega^2}\,\D\tau\,\D\chi
\nonumber\\
& 
{}-\gamma^2\left(1+\frac{g^2}{\Omega^2}-v^2\right)\D\chi^2
\nonumber\\
&
{}-\left(1+\frac{g^2}{\Omega^2}\right)\D y^2 
\text{ .}
\end{align}
Since $1/\clow^2=1+g^2/\Omega^2$ according to Eq.~\eqref{eq:c-low-energy}, 
this corresponds to Eq.~\eqref{g_00}.
Calculating the surface gravity from the above metric,
\begin{equation}
\label{eq:surface-gravity}
\kappa
=
\frac{1}{2}\left|\frac{\partial_\chi \effmetric_{00}}{\effmetric_{01}}\right|
\Biggr|_{\mathrm{horizon}}
=
v\xi 
\text{ ,}
\end{equation}
we find that the Hawking temperature is given by the standard 
expression (as expected)
\begin{equation}
\THawking=\frac{\kappa}{2\pi}=\frac{v\xi}{2\pi}
\text{ .}
\end{equation}
Note that the transformation from the above stationary 
Painlev{\'e}--Gullstrand--Lema{\^\i}tre-type coordinates $\tau$ and $\chi$
to static Schwarzschild-type coordinates $\tau_\ast$ and $\chi$ via 
\begin{equation}
\D \tau_\ast = \D \tau + \frac{\effmetric_{01}}{\effmetric_{00}}\,\D\chi
\end{equation}
does not change the $\effmetric_{00}$ component of the metric,
\begin{align}
\D\proptimeeff^2={}
&
\gamma^2\left(1-v^2\left[1+\frac{g^2}{\Omega^2}\right]\right)\D\tau_\ast^2
\nonumber\\
&
{}-\gamma^{-2}\left(1-v^2\left[1+\frac{g^2}{\Omega^2}\right]\right)^{-1}
\left(1+\frac{g^2}{\Omega^2}\right)\D\chi^2
\nonumber\\
&
{}-\left(1+\frac{g^2}{\Omega^2}\right)\D y^2
\text{ .}
\end{align}
In complete analogy to the Schwarzschild metric, we may introduce the 
tortoise coordinate $\chi_\ast$ via
\begin{equation}
\D\chi_\ast
=
\gamma^{-2}\left(1-v^2\left[1+\frac{g^2}{\Omega^2}\right]\right)^{-1}
\sqrt{1+\frac{g^2}{\Omega^2}}\,\D\chi\text{ ,}
\end{equation}
such that the metric becomes
\begin{align}
\D\proptimeeff^2={}
&
\gamma^2\left(1-v^2\left[1+\frac{g^2}{\Omega^2}\right]\right)
\left(\D\tau_\ast^2-\D\chi_\ast^2\right)
\nonumber\\
&
{}-\left(1+\frac{g^2}{\Omega^2}\right)\D y^2
\text{ .}
\end{align}
For a large $\chi_\ast\to+\infty$, this coordinate coincides with the original
one, $\chi_\ast\approx\chi$, while the other limit, $\chi_\ast\to-\infty$,
approaches the horizon, $\chi\downarrow0$.

%%%%%%%%%%%%%%%%%%%%%%%%%%%%%%%%%%%%%%%%%%%%%%%%%%%%%%%%%%%%%%%%%%%%%%%%%%%%%%%%
\subsection{Breakdown of conformal invariance}
%%%%%%%%%%%%%%%%%%%%%%%%%%%%%%%%%%%%%%%%%%%%%%%%%%%%%%%%%%%%%%%%%%%%%%%%%%%%%%%%

The fact that the effective action~\eqref{eq:Leff} is not conformally 
invariant has important consequences.
One of them is that left- and right-mov\-ing modes are not decoupled 
from each other.
In our case, we find that the counterpropagating mode couples to 
the other modes (such as the Hawking radiation), which results in 
the second branch cut and the additional Bogoliubov coefficient 
$\eta_\omega$, which in turn gives rise to the gray-body factor 
$\GBF_\omega$.

If we consider the limit where $v$ approaches $1$ 
(the vacuum speed of light), we find that $g^2/\Omega^2$ 
becomes very small (near the horizon) and thus the effective 
action~\eqref{eq:Leff} is nearly conformal 
(even though one has to be careful with such an asymptotic statement).
In this limit, the counterpropagating mode decouples approximately.
Therefore, the second branch-cut contribution ($\psicutm_{\omega}$) and the
associated additional Bogoliubov coefficient $\eta_\omega$ go to zero such that
the gray-body factor approaches unity and the Hawking temperature converges to
the ordinary expression $\xi/(2\pi)$.
As one can easily imagine, a pulse with a very small polarizability 
moving with almost the vacuum speed of light has very little impact 
on counterpropagating photons.

Note that the effective action~\eqref{eq:Leff} could be made 
conformally invariant if we added a magnetic permeability $\mu$  
to the dielectric permittivity $\varepsilon$ and demanded that 
$\varepsilon=\mu$.
In this case,
$\Leff=(\varepsilon|\partial_{t}A|^{2}-|\partial_{x}A|^{2}/\mu)/2$,
we may introduce an effective metric such as
$\D\proptimeeff^2=\D t^2-\varepsilon^2\D x^2$, which is 1+1 dimensional
and thus can be cast into a conformally flat form.
This conformal invariance leads to several simplifications (e.g., decoupling of
left- and right-mov\-ing modes), which have been exploited in the literature;
see, e.g., Refs.~\cite{Brout:Sonic-dispersion,Schuetzhold+Unruh:Particle-origin,
Coutant+al:S-matrix-approach}.
How\-ever---as we have seen above---these simplifications are not
necessary for our purpose (the derivation of Hawking radiation).
Of course, they can make not only obtaining but also interpreting 
the results easier.
For example, as we discuss below in Sec.~\ref{Conclusions},
identifying the correct Hawking temperature in the case of broken conformal
symmetry requires more care than in the conformally invariant situation, where
one can easily read it off the Hawking spectrum.

%%%%%%%%%%%%%%%%%%%%%%%%%%%%%%%%%%%%%%%%%%%%%%%%%%%%%%%%%%%%%%%%%%%%%%%%%%%%%%%%
\subsection{Effective potential}
%%%%%%%%%%%%%%%%%%%%%%%%%%%%%%%%%%%%%%%%%%%%%%%%%%%%%%%%%%%%%%%%%%%%%%%%%%%%%%%%

\begin{figure}
\includegraphics{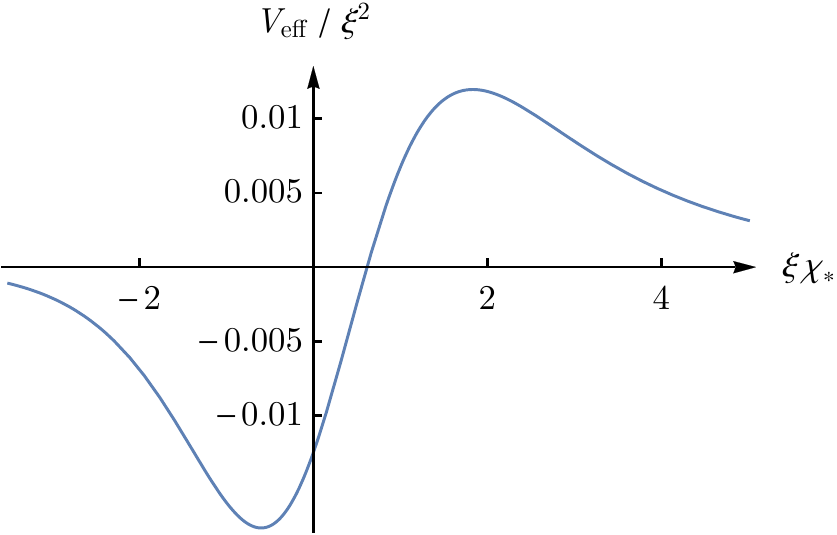}
\caption{\label{fig:Veff}%
Effective potential~\protect\eqref{eq:Veff} outside the black hole as a function
of the tortoise coordinate $\chi_\ast$ for $v=2/3$ (however, the shape of the
graph is generic for arbitrary pulse velocities). The coordinate transformation
$\chi=\chi(\chi_\ast)$ was carried out numerically.%
}
\end{figure}

The scattering of modes (coupling between left- and right-mov\-ing modes)
due to the breakdown of conformal invariance can be understood nicely
in terms of the effective potential.
After transforming to the aforementioned tortoise coordinate $\chi_\ast$ 
and rescaling $A_\omega$ according to 
$\mathcal{A}_\omega=|\effmetric_{22}|^{1/4} A_\omega$, the wave equation
reads
\begin{equation}
\label{eq:effective-potential}
\left[\omega^2+\frac{\partial^2}{\partial\chi_\ast^2}-\Veff(\chi_\ast)\right]
\mathcal{A}_\omega=0\text{ ,}
\end{equation}
with the effective potential
\begin{equation}
\Veff
=
-v^2\xi^3\chi\,(1-v^2)\,
\frac{2-(3+v^2)\xi\chi-16v^2\xi^2\chi^2}{(1+2\xi\chi)^3(1+2v^2\xi\chi)^3}
\text{ .}
\label{eq:Veff}
\end{equation}
For large $\chi_\ast\to+\infty$, this potential decreases as $1/\chi^3_\ast$
while in the other limit, $\chi_\ast\to-\infty$, it decreases exponentially
(when approaching the horizon); see Fig.~\ref{fig:Veff}.

Note that Eq.~\eqref{eq:effective-potential} is formally equivalent to a
Schr\"odinger scattering problem with the potential $\Veff$ and the
nonrelativistic energy $\energy\propto\omega^2$.
Thus, we get the usual transmission and reflection coefficients.

%%%%%%%%%%%%%%%%%%%%%%%%%%%%%%%%%%%%%%%%%%%%%%%%%%%%%%%%%%%%%%%%%%%%%%%%%%%%%%%%
%%%%%%%%%%%%%%%%%%%%%%%%%%%%%%%%%%%%%%%%%%%%%%%%%%%%%%%%%%%%%%%%%%%%%%%%%%%%%%%%

%%%%%%%%%%%%%%%%%%%%%%%%%%%%%%%%%%%%%%%%%%%%%%%%%%%%%%%%%%%%%%%%%%%%%%%%%%%%%%%%
%%%%%%%%%%%%%%%%%%%%%%%%%%%%%%%%%%%%%%%%%%%%%%%%%%%%%%%%%%%%%%%%%%%%%%%%%%%%%%%%
\section{\label{Conclusions}Conclusions}
%%%%%%%%%%%%%%%%%%%%%%%%%%%%%%%%%%%%%%%%%%%%%%%%%%%%%%%%%%%%%%%%%%%%%%%%%%%%%%%%
%%%%%%%%%%%%%%%%%%%%%%%%%%%%%%%%%%%%%%%%%%%%%%%%%%%%%%%%%%%%%%%%%%%%%%%%%%%%%%%%

Based on the Hopfield model~\eqref{eq:Llab}, we presented a fully relativistic 
derivation of the analog of Hawking radiation in a dispersive dielectric medium
employing a minimal set of as\-sump\-tions/ap\-prox\-i\-ma\-tions.
As expected, we find that the Hawking temperature~\eqref{eq:Hawking-temp-pls} is
set by the surface gravity~\eqref{eq:surface-gravity}, but we also obtain a
gray-body factor~\eqref{eq:gray-body-factor-pls}, which results in a nonthermal
spectrum~\eqref{eq:Hawking-spectrum-pls} observed at infinity.

Let us emphasize that just this nonthermal spectrum
$\braket{\nH_\omega}_\initial=|\beta_{\omega}|^2$ is not enough to determine the
Hawking temperature.
Instead, the Hawking temperature can be read off the ratio~\eqref{ratio} of the
Bogoliubov coefficients.
This identification can be based on the following picture:
imagine an initial short-wave\-length wave packet composed of $\kstarplus$ and
$\kstarminus$ modes with this ratio~\eqref{ratio}.
Due to its short wavelength (and its thereby reduced group velocity), this
wave packet approaches the horizon without scattering (by the effective
potential $\Veff$) and is constantly stretched during this approach (analog
of gravitational redshift).
Near the horizon, it is then stretched to long wavelengths and starts to 
escape to infinity as Hawking radiation (as its group velocity has increased).
However, on the way out to infinity, part of this Hawking wave packet is 
scattered (by the effective potential $\Veff$ due to the breaking of 
conformal invariance) and thereby transformed into a counterpropagating 
mode, which is then swallowed by the horizon.
Now, in order to have only the Hawking mode at late times, we can send in an
additional counterpropagating mode such that the two amplitudes (this direct
counterpropagating mode and the scattered mode) beyond the horizon precisely
cancel each other.
This then leads to the form~\eqref{eq:Bogoliubov-trafo}, which relates the
final Hawking mode to a linear combination of initial short-wave\-length 
modes (with amplitudes $\alpha_{\omega}$ and $\beta_{\omega}$) plus 
an initial counterpropagating mode (with the amplitude $\eta_{\omega}$).
In view of this scale separation (short and long wavelengths), we may
distinguish the pure scattering process (leading to $\eta_\omega$), which does
not mix pos\-i\-tive- and neg\-a\-tive-norm states, from the mechanism of
particle creation (involving the amplitudes $\alpha_\omega$ and $\beta_\omega$).
The latter can be understood as an amplification process due to the horizon 
and is associated with the Hawking temperature, while the former gives rise to 
the gray-body factor.

As another way of demonstrating that the ratio~\eqref{ratio} determines the 
Hawking temperature, let us reconstruct the analog of the
Israel--Hartle--Hawking state (thermal equilibrium) and imagine sending in the
long-wave\-length modes in a thermal state with the temperature $\TIHH$, while
the short-wave\-length modes are still in their vacuum state.
In this situation, the expectation value for the Hawking particles reads
\begin{align}
\Braket{\nH_\omega}_\initial
&
=|\beta_\omega|^2+|\eta_\omega|^2 \Braket{\ncp_\omega}_\initial
\nonumber\\
&
=\frac{1-|\eta_\omega|^2}{\E^{\omega/\THawking}-1}
+
\frac{|\eta_\omega|^2}{\E^{\omega/\TIHH}-1}\text{ ,}
\end{align}
where we have inserted the above result~\eqref{eq:Hawking-spectrum-pls}
for $|\beta_\omega|^2$ and the Bose--Einstein distribution (with temperature
$\TIHH$) for $\braket{\ncp_\omega}_\initial$.
We see that for $\TIHH=\THawking$, we obtain a thermal spectrum for
$\braket{\nH_\omega}_\initial$ with that temperature---which precisely
corresponds to the detailed balance condition, as expected in a thermal
equilibrium state.

Note that for obtaining the above results---such as the Hawking temperature 
as determined by the surface gravity and the gray-body factor---it is essential 
to take both branch cuts and thus also the counterpropagating mode into account.
This might be one reason for the difference between our results and those in the
recent paper~\cite{Belgiorno:Hopfield-model-results}.
In their paper, the second branch cut and the counterpropagating mode are 
apparently neglected and the Hawking temperature obtained there differs 
from our expression (and thus also from the surface gravity).

In our derivation, we had both branch cuts running upwards in the complex plane
because this was most convenient for obtaining the solution which is evanescent
inside the black hole; see the discussion of the boundary condition in 
Sec.~\ref{sub:JWKB-modes-inside}.
For other choices of the branch cuts (with respect to the integration contour),
we would obtain solutions with different boundary conditions.
For example, having the two branch cuts run downwards to infinity 
(e.g., one in the bottom-right valley and the other one in the bottom-left
valley), an analogous calculation would give the solution with the two initial
short-wave\-length solutions, $\kstarplusminus$, outside and the two final
modes, $\ktildestarp$ and $\ktildestarcp$, inside the black hole.
In this way, by considering all possible combinations (both branch cuts up or
both down or one up and one down), one can derive the 3$\times$3 ``scattering''
matrix which connects the three initial modes, $\kstarplusminus$ and $\kstarcp$
(all outside), with the three final modes, $\ktildestarp$ and $\ktildestarcp$ 
(inside), as well as $\kstarH$ (outside).

%%%%%%%%%%%%%%%%%%%%%%%%%%%%%%%%%%%%%%%%%%%%%%%%%%%%%%%%%%%%%%%%%%%%%%%%%%%%%%%%
\begin{acknowledgments}
%%%%%%%%%%%%%%%%%%%%%%%%%%%%%%%%%%%%%%%%%%%%%%%%%%%%%%%%%%%%%%%%%%%%%%%%%%%%%%%%

W.G.U.\ obtained support from Natural Sciences and Engineering Research Council
of Canada (NSERC), the Templeton Foundation, and the Canadian Institute for
Advanced Research.
R.S.\ acknowledges support from Deutsche Forschungsgemeinschaft (SFB-TR12).
We thank the Perimeter Institute for Theoretical Physics (PI),
the Mainz Institute for Theoretical Physics (MITP), and the
Pacific Institute of Theoretical Physics (PITP), where part of this work was
done, for their hospitality and support.
R.S.\ also wishes to thank the
Frankfurt Institute for Advanced Studies (FIAS) for the hospitality and support.

%%%%%%%%%%%%%%%%%%%%%%%%%%%%%%%%%%%%%%%%%%%%%%%%%%%%%%%%%%%%%%%%%%%%%%%%%%%%%%%%
\end{acknowledgments}
%%%%%%%%%%%%%%%%%%%%%%%%%%%%%%%%%%%%%%%%%%%%%%%%%%%%%%%%%%%%%%%%%%%%%%%%%%%%%%%%

%\bibliography{references}

\begin{thebibliography}{61}%
\makeatletter
\providecommand \@ifxundefined [1]{%
 \@ifx{#1\undefined}
}%
\providecommand \@ifnum [1]{%
 \ifnum #1\expandafter \@firstoftwo
 \else \expandafter \@secondoftwo
 \fi
}%
\providecommand \@ifx [1]{%
 \ifx #1\expandafter \@firstoftwo
 \else \expandafter \@secondoftwo
 \fi
}%
\providecommand \natexlab [1]{#1}%
\providecommand \enquote  [1]{``#1''}%
\providecommand \bibnamefont  [1]{#1}%
\providecommand \bibfnamefont [1]{#1}%
\providecommand \citenamefont [1]{#1}%
\providecommand \href@noop [0]{\@secondoftwo}%
\providecommand \href [0]{\begingroup \@sanitize@url \@href}%
\providecommand \@href[1]{\@@startlink{#1}\@@href}%
\providecommand \@@href[1]{\endgroup#1\@@endlink}%
\providecommand \@sanitize@url [0]{\catcode `\\12\catcode `\$12\catcode
  `\&12\catcode `\#12\catcode `\^12\catcode `\_12\catcode `\%12\relax}%
\providecommand \@@startlink[1]{}%
\providecommand \@@endlink[0]{}%
\providecommand \url  [0]{\begingroup\@sanitize@url \@url }%
\providecommand \@url [1]{\endgroup\@href {#1}{\urlprefix }}%
\providecommand \urlprefix  [0]{URL }%
\providecommand \Eprint [0]{\href }%
\providecommand \doibase [0]{http://dx.doi.org/}%
\providecommand \selectlanguage [0]{\@gobble}%
\providecommand \bibinfo  [0]{\@secondoftwo}%
\providecommand \bibfield  [0]{\@secondoftwo}%
\providecommand \translation [1]{[#1]}%
\providecommand \BibitemOpen [0]{}%
\providecommand \bibitemStop [0]{}%
\providecommand \bibitemNoStop [0]{.\EOS\space}%
\providecommand \EOS [0]{\spacefactor3000\relax}%
\providecommand \BibitemShut  [1]{\csname bibitem#1\endcsname}%
\let\auto@bib@innerbib\@empty
%</preamble>
\bibitem [{\citenamefont {Hawking}(1974)}]{Hawking:1}%
  \BibitemOpen
  \bibfield  {author} {\bibinfo {author} {\bibfnamefont {S.~W.}\ \bibnamefont
  {Hawking}},\ }\bibfield  {title} {\enquote {\bibinfo {title} {Black hole
  explosions?}}\ }\href {\doibase 10.1038/248030a0} {\bibfield  {journal}
  {\bibinfo  {journal} {Nature}\ }\textbf {\bibinfo {volume} {248}},\ \bibinfo
  {pages} {30--31} (\bibinfo {year} {1974})}\BibitemShut {NoStop}%
\bibitem [{\citenamefont {Hawking}(1975)}]{Hawking:2}%
  \BibitemOpen
  \bibfield  {author} {\bibinfo {author} {\bibfnamefont {S.~W.}\ \bibnamefont
  {Hawking}},\ }\bibfield  {title} {\enquote {\bibinfo {title} {Particle
  creation by black holes},}\ }\href {\doibase 10.1007/BF02345020} {\bibfield
  {journal} {\bibinfo  {journal} {Commun. math. Phys.}\ }\textbf {\bibinfo
  {volume} {43}},\ \bibinfo {pages} {199--220} (\bibinfo {year}
  {1975})}\BibitemShut {NoStop}%
\bibitem [{\citenamefont {Unruh}(1981)}]{Unruh:Sonic-analog}%
  \BibitemOpen
  \bibfield  {author} {\bibinfo {author} {\bibfnamefont {W.~G.}\ \bibnamefont
  {Unruh}},\ }\bibfield  {title} {\enquote {\bibinfo {title} {Experimental
  {B}lack-{H}ole {E}vaporation?}}\ }\href {\doibase
  10.1103/PhysRevLett.46.1351} {\bibfield  {journal} {\bibinfo  {journal}
  {Phys. Rev. Lett.}\ }\textbf {\bibinfo {volume} {46}},\ \bibinfo {pages}
  {1351--1353} (\bibinfo {year} {1981})}\BibitemShut {NoStop}%
\bibitem [{\citenamefont {Unruh}(1995)}]{Unruh:Sonic-dispersion}%
  \BibitemOpen
  \bibfield  {author} {\bibinfo {author} {\bibfnamefont {W.~G.}\ \bibnamefont
  {Unruh}},\ }\bibfield  {title} {\enquote {\bibinfo {title} {Sonic analogue of
  black holes and the effects of high frequencies on black hole evaporation},}\
  }\href {\doibase 10.1103/PhysRevD.51.2827} {\bibfield  {journal} {\bibinfo
  {journal} {Phys. Rev. D}\ }\textbf {\bibinfo {volume} {51}},\ \bibinfo
  {pages} {2827--2838} (\bibinfo {year} {1995})}\BibitemShut {NoStop}%
\bibitem [{\citenamefont {Brout}\ \emph {et~al.}(1995)\citenamefont {Brout},
  \citenamefont {Massar}, \citenamefont {Parentani},\ and\ \citenamefont
  {Spindel}}]{Brout:Sonic-dispersion}%
  \BibitemOpen
  \bibfield  {author} {\bibinfo {author} {\bibfnamefont {R.}~\bibnamefont
  {Brout}}, \bibinfo {author} {\bibfnamefont {S.}~\bibnamefont {Massar}},
  \bibinfo {author} {\bibfnamefont {R.}~\bibnamefont {Parentani}}, \ and\
  \bibinfo {author} {\bibfnamefont {Ph.}\ \bibnamefont {Spindel}},\ }\bibfield
  {title} {\enquote {\bibinfo {title} {Hawking radiation without
  trans-{P}lanckian frequencies},}\ }\href {\doibase 10.1103/PhysRevD.52.4559}
  {\bibfield  {journal} {\bibinfo  {journal} {Phys. Rev. D}\ }\textbf {\bibinfo
  {volume} {52}},\ \bibinfo {pages} {4559--4568} (\bibinfo {year}
  {1995})}\BibitemShut {NoStop}%
\bibitem [{\citenamefont {Jacobson}(1996)}]{Jacobson:Origin}%
  \BibitemOpen
  \bibfield  {author} {\bibinfo {author} {\bibfnamefont {T.}~\bibnamefont
  {Jacobson}},\ }\bibfield  {title} {\enquote {\bibinfo {title} {On the origin
  of the outgoing black hole modes},}\ }\href {\doibase
  10.1103/PhysRevD.53.7082} {\bibfield  {journal} {\bibinfo  {journal} {Phys.
  Rev. D}\ }\textbf {\bibinfo {volume} {53}},\ \bibinfo {pages} {7082--7088}
  (\bibinfo {year} {1996})}\BibitemShut {NoStop}%
\bibitem [{\citenamefont {Corley}\ and\ \citenamefont
  {Jacobson}(1996)}]{Corley:Sonic-dispersion}%
  \BibitemOpen
  \bibfield  {author} {\bibinfo {author} {\bibfnamefont {S.}~\bibnamefont
  {Corley}}\ and\ \bibinfo {author} {\bibfnamefont {T.}~\bibnamefont
  {Jacobson}},\ }\bibfield  {title} {\enquote {\bibinfo {title} {Hawking
  spectrum and high frequency dispersion},}\ }\href {\doibase
  10.1103/PhysRevD.54.1568} {\bibfield  {journal} {\bibinfo  {journal} {Phys.
  Rev. D}\ }\textbf {\bibinfo {volume} {54}},\ \bibinfo {pages} {1568--1586}
  (\bibinfo {year} {1996})}\BibitemShut {NoStop}%
\bibitem [{\citenamefont {Visser}(1998)}]{Visser:Sonic-analog}%
  \BibitemOpen
  \bibfield  {author} {\bibinfo {author} {\bibfnamefont {M.}~\bibnamefont
  {Visser}},\ }\bibfield  {title} {\enquote {\bibinfo {title} {Acoustic black
  holes: horizons, ergospheres and {H}awking radiation},}\ }\href {\doibase
  10.1088/0264-9381/15/6/024} {\bibfield  {journal} {\bibinfo  {journal}
  {Class. Quantum Grav.}\ }\textbf {\bibinfo {volume} {15}},\ \bibinfo {pages}
  {1767--1791} (\bibinfo {year} {1998})}\BibitemShut {NoStop}%
\bibitem [{\citenamefont {Corley}(1998)}]{Corley:Spectrum-calculation}%
  \BibitemOpen
  \bibfield  {author} {\bibinfo {author} {\bibfnamefont {S.}~\bibnamefont
  {Corley}},\ }\bibfield  {title} {\enquote {\bibinfo {title} {Computing the
  spectrum of black hole radiation in the presence of high frequency
  dispersion: {A}n analytical approach},}\ }\href {\doibase
  10.1103/PhysRevD.57.6280} {\bibfield  {journal} {\bibinfo  {journal} {Phys.
  Rev. D}\ }\textbf {\bibinfo {volume} {57}},\ \bibinfo {pages} {6280--6291}
  (\bibinfo {year} {1998})}\BibitemShut {NoStop}%
\bibitem [{\citenamefont {Himemoto}\ and\ \citenamefont
  {Tanaka}(2000)}]{Yoshiaki+Takahiro}%
  \BibitemOpen
  \bibfield  {author} {\bibinfo {author} {\bibfnamefont {Y.}~\bibnamefont
  {Himemoto}}\ and\ \bibinfo {author} {\bibfnamefont {T.}~\bibnamefont
  {Tanaka}},\ }\bibfield  {title} {\enquote {\bibinfo {title} {Generalization
  of the model of {H}awking radiation with modified high frequency dispersion
  relation},}\ }\href {\doibase 10.1103/PhysRevD.61.064004} {\bibfield
  {journal} {\bibinfo  {journal} {Phys. Rev. D}\ }\textbf {\bibinfo {volume}
  {61}},\ \bibinfo {pages} {064004} (\bibinfo {year} {2000})}\BibitemShut
  {NoStop}%
\bibitem [{\citenamefont {Saida}\ and\ \citenamefont
  {Sakagami}(2000)}]{Saida+Sakagami}%
  \BibitemOpen
  \bibfield  {author} {\bibinfo {author} {\bibfnamefont {H.}~\bibnamefont
  {Saida}}\ and\ \bibinfo {author} {\bibfnamefont {M.}~\bibnamefont
  {Sakagami}},\ }\bibfield  {title} {\enquote {\bibinfo {title} {Black hole
  radiation with high frequency dispersion},}\ }\href {\doibase
  10.1103/PhysRevD.61.084023} {\bibfield  {journal} {\bibinfo  {journal} {Phys.
  Rev. D}\ }\textbf {\bibinfo {volume} {61}},\ \bibinfo {pages} {084023}
  (\bibinfo {year} {2000})}\BibitemShut {NoStop}%
\bibitem [{\citenamefont {Unruh}\ and\ \citenamefont
  {Sch\"utzhold}(2005)}]{Unruh+Schuetzhold:Universality}%
  \BibitemOpen
  \bibfield  {author} {\bibinfo {author} {\bibfnamefont {W.~G.}\ \bibnamefont
  {Unruh}}\ and\ \bibinfo {author} {\bibfnamefont {R.}~\bibnamefont
  {Sch\"utzhold}},\ }\bibfield  {title} {\enquote {\bibinfo {title}
  {Universality of the {H}awking effect},}\ }\href {\doibase
  10.1103/PhysRevD.71.024028} {\bibfield  {journal} {\bibinfo  {journal} {Phys.
  Rev. D}\ }\textbf {\bibinfo {volume} {71}},\ \bibinfo {pages} {024028}
  (\bibinfo {year} {2005})}\BibitemShut {NoStop}%
\bibitem [{\citenamefont {Agull\'o}\ \emph {et~al.}(2007)\citenamefont
  {Agull\'o}, \citenamefont {Navarro-Salas}, \citenamefont {Olmo},\ and\
  \citenamefont {Parker}}]{Agullo-et-al}%
  \BibitemOpen
  \bibfield  {author} {\bibinfo {author} {\bibfnamefont {I.}~\bibnamefont
  {Agull\'o}}, \bibinfo {author} {\bibfnamefont {J.}~\bibnamefont
  {Navarro-Salas}}, \bibinfo {author} {\bibfnamefont {G.~J.}\ \bibnamefont
  {Olmo}}, \ and\ \bibinfo {author} {\bibfnamefont {L.}~\bibnamefont
  {Parker}},\ }\bibfield  {title} {\enquote {\bibinfo {title} {Short-distance
  contribution to the spectrum of {H}awking radiation},}\ }\href {\doibase
  10.1103/PhysRevD.76.044018} {\bibfield  {journal} {\bibinfo  {journal} {Phys.
  Rev. D}\ }\textbf {\bibinfo {volume} {76}},\ \bibinfo {pages} {044018}
  (\bibinfo {year} {2007})}\BibitemShut {NoStop}%
\bibitem [{\citenamefont {Sch\"utzhold}\ and\ \citenamefont
  {Unruh}(2008)}]{Schuetzhold+Unruh:Particle-origin}%
  \BibitemOpen
  \bibfield  {author} {\bibinfo {author} {\bibfnamefont {R.}~\bibnamefont
  {Sch\"utzhold}}\ and\ \bibinfo {author} {\bibfnamefont {W.~G.}\ \bibnamefont
  {Unruh}},\ }\bibfield  {title} {\enquote {\bibinfo {title} {Origin of the
  particles in black hole evaporation},}\ }\href {\doibase
  10.1103/PhysRevD.78.041504} {\bibfield  {journal} {\bibinfo  {journal} {Phys.
  Rev. D}\ }\textbf {\bibinfo {volume} {78}},\ \bibinfo {pages} {041504}
  (\bibinfo {year} {2008})}\BibitemShut {NoStop}%
\bibitem [{\citenamefont {Macher}\ and\ \citenamefont
  {Parentani}(2009{\natexlab{a}})}]{Macher+Parentani:BWH-radiation}%
  \BibitemOpen
  \bibfield  {author} {\bibinfo {author} {\bibfnamefont {J.}~\bibnamefont
  {Macher}}\ and\ \bibinfo {author} {\bibfnamefont {R.}~\bibnamefont
  {Parentani}},\ }\bibfield  {title} {\enquote {\bibinfo {title} {Black/white
  hole radiation from dispersive theories},}\ }\href {\doibase
  10.1103/PhysRevD.79.124008} {\bibfield  {journal} {\bibinfo  {journal} {Phys.
  Rev. D}\ }\textbf {\bibinfo {volume} {79}},\ \bibinfo {pages} {124008}
  (\bibinfo {year} {2009}{\natexlab{a}})}\BibitemShut {NoStop}%
\bibitem [{\citenamefont {Finazzi}\ and\ \citenamefont
  {Parentani}(2011)}]{Finazzi+Parentani:Broadening}%
  \BibitemOpen
  \bibfield  {author} {\bibinfo {author} {\bibfnamefont {S.}~\bibnamefont
  {Finazzi}}\ and\ \bibinfo {author} {\bibfnamefont {R.}~\bibnamefont
  {Parentani}},\ }\bibfield  {title} {\enquote {\bibinfo {title} {Spectral
  properties of acoustic black hole radiation: {B}roadening the horizon},}\
  }\href {\doibase 10.1103/PhysRevD.83.084010} {\bibfield  {journal} {\bibinfo
  {journal} {Phys. Rev. D}\ }\textbf {\bibinfo {volume} {83}},\ \bibinfo
  {pages} {084010} (\bibinfo {year} {2011})}\BibitemShut {NoStop}%
\bibitem [{\citenamefont {Coutant}\ \emph {et~al.}(2012)\citenamefont
  {Coutant}, \citenamefont {Parentani},\ and\ \citenamefont
  {Finazzi}}]{Coutant+al:S-matrix-approach}%
  \BibitemOpen
  \bibfield  {author} {\bibinfo {author} {\bibfnamefont {A.}~\bibnamefont
  {Coutant}}, \bibinfo {author} {\bibfnamefont {R.}~\bibnamefont {Parentani}},
  \ and\ \bibinfo {author} {\bibfnamefont {S.}~\bibnamefont {Finazzi}},\
  }\bibfield  {title} {\enquote {\bibinfo {title} {Black hole radiation with
  short distance dispersion, an analytical {$S$}-matrix approach},}\ }\href
  {\doibase 10.1103/PhysRevD.85.024021} {\bibfield  {journal} {\bibinfo
  {journal} {Phys. Rev. D}\ }\textbf {\bibinfo {volume} {85}},\ \bibinfo
  {pages} {024021} (\bibinfo {year} {2012})}\BibitemShut {NoStop}%
\bibitem [{\citenamefont {Finazzi}\ and\ \citenamefont
  {Parentani}(2012)}]{Finazzi+Parentani:Two-regimes}%
  \BibitemOpen
  \bibfield  {author} {\bibinfo {author} {\bibfnamefont {S.}~\bibnamefont
  {Finazzi}}\ and\ \bibinfo {author} {\bibfnamefont {R.}~\bibnamefont
  {Parentani}},\ }\bibfield  {title} {\enquote {\bibinfo {title} {Hawking
  radiation in dispersive theories, the two regimes},}\ }\href {\doibase
  10.1103/PhysRevD.85.124027} {\bibfield  {journal} {\bibinfo  {journal} {Phys.
  Rev. D}\ }\textbf {\bibinfo {volume} {85}},\ \bibinfo {pages} {124027}
  (\bibinfo {year} {2012})}\BibitemShut {NoStop}%
\bibitem [{\citenamefont {Coutant}\ and\ \citenamefont
  {Parentani}(2014)}]{Coutant+Parentani:Paradigm}%
  \BibitemOpen
  \bibfield  {author} {\bibinfo {author} {\bibfnamefont {A.}~\bibnamefont
  {Coutant}}\ and\ \bibinfo {author} {\bibfnamefont {R.}~\bibnamefont
  {Parentani}},\ }\bibfield  {title} {\enquote {\bibinfo {title} {Hawking
  radiation with dispersion: {T}he broadened horizon paradigm},}\ }\href
  {\doibase 10.1103/PhysRevD.90.121501} {\bibfield  {journal} {\bibinfo
  {journal} {Phys. Rev. D}\ }\textbf {\bibinfo {volume} {90}},\ \bibinfo
  {pages} {121501} (\bibinfo {year} {2014})}\BibitemShut {NoStop}%
\bibitem [{\citenamefont {Garay}\ \emph {et~al.}(2000)\citenamefont {Garay},
  \citenamefont {Anglin}, \citenamefont {Cirac},\ and\ \citenamefont
  {Zoller}}]{BEC:Analog-1}%
  \BibitemOpen
  \bibfield  {author} {\bibinfo {author} {\bibfnamefont {L.~J.}\ \bibnamefont
  {Garay}}, \bibinfo {author} {\bibfnamefont {J.~R.}\ \bibnamefont {Anglin}},
  \bibinfo {author} {\bibfnamefont {J.~I.}\ \bibnamefont {Cirac}}, \ and\
  \bibinfo {author} {\bibfnamefont {P.}~\bibnamefont {Zoller}},\ }\bibfield
  {title} {\enquote {\bibinfo {title} {Sonic analog of {G}ravitational {B}lack
  {H}oles in {B}ose-{E}instein {C}ondensates},}\ }\href {\doibase
  10.1103/PhysRevLett.85.4643} {\bibfield  {journal} {\bibinfo  {journal}
  {Phys. Rev. Lett.}\ }\textbf {\bibinfo {volume} {85}},\ \bibinfo {pages}
  {4643--4647} (\bibinfo {year} {2000})}\BibitemShut {NoStop}%
\bibitem [{\citenamefont {Garay}\ \emph {et~al.}(2001)\citenamefont {Garay},
  \citenamefont {Anglin}, \citenamefont {Cirac},\ and\ \citenamefont
  {Zoller}}]{BEC:Analog-2}%
  \BibitemOpen
  \bibfield  {author} {\bibinfo {author} {\bibfnamefont {L.~J.}\ \bibnamefont
  {Garay}}, \bibinfo {author} {\bibfnamefont {J.~R.}\ \bibnamefont {Anglin}},
  \bibinfo {author} {\bibfnamefont {J.~I.}\ \bibnamefont {Cirac}}, \ and\
  \bibinfo {author} {\bibfnamefont {P.}~\bibnamefont {Zoller}},\ }\bibfield
  {title} {\enquote {\bibinfo {title} {Sonic black holes in dilute
  {B}ose-{E}instein condensates},}\ }\href {\doibase
  10.1103/PhysRevA.63.023611} {\bibfield  {journal} {\bibinfo  {journal} {Phys.
  Rev. A}\ }\textbf {\bibinfo {volume} {63}},\ \bibinfo {pages} {023611}
  (\bibinfo {year} {2001})}\BibitemShut {NoStop}%
\bibitem [{\citenamefont {Macher}\ and\ \citenamefont
  {Parentani}(2009{\natexlab{b}})}]{BEC:Analog-3}%
  \BibitemOpen
  \bibfield  {author} {\bibinfo {author} {\bibfnamefont {J.}~\bibnamefont
  {Macher}}\ and\ \bibinfo {author} {\bibfnamefont {R.}~\bibnamefont
  {Parentani}},\ }\bibfield  {title} {\enquote {\bibinfo {title} {Black-hole
  radiation in {B}ose-{E}instein condensates},}\ }\href {\doibase
  10.1103/PhysRevA.80.043601} {\bibfield  {journal} {\bibinfo  {journal} {Phys.
  Rev. A}\ }\textbf {\bibinfo {volume} {80}},\ \bibinfo {pages} {043601}
  (\bibinfo {year} {2009}{\natexlab{b}})}\BibitemShut {NoStop}%
\bibitem [{\citenamefont {Lahav}\ \emph {et~al.}(2010)\citenamefont {Lahav},
  \citenamefont {Itah}, \citenamefont {Blumkin}, \citenamefont {Gordon},
  \citenamefont {Rinott}, \citenamefont {Zayats},\ and\ \citenamefont
  {Steinhauer}}]{Steinhauer:1}%
  \BibitemOpen
  \bibfield  {author} {\bibinfo {author} {\bibfnamefont {O.}~\bibnamefont
  {Lahav}}, \bibinfo {author} {\bibfnamefont {A.}~\bibnamefont {Itah}},
  \bibinfo {author} {\bibfnamefont {A.}~\bibnamefont {Blumkin}}, \bibinfo
  {author} {\bibfnamefont {C.}~\bibnamefont {Gordon}}, \bibinfo {author}
  {\bibfnamefont {S.}~\bibnamefont {Rinott}}, \bibinfo {author} {\bibfnamefont
  {A.}~\bibnamefont {Zayats}}, \ and\ \bibinfo {author} {\bibfnamefont
  {J.}~\bibnamefont {Steinhauer}},\ }\bibfield  {title} {\enquote {\bibinfo
  {title} {Realization of a {S}onic {B}lack {H}ole {A}nalog in a
  {B}ose-{E}instein {C}ondensate},}\ }\href {\doibase
  10.1103/PhysRevLett.105.240401} {\bibfield  {journal} {\bibinfo  {journal}
  {Phys. Rev. Lett.}\ }\textbf {\bibinfo {volume} {105}},\ \bibinfo {pages}
  {240401} (\bibinfo {year} {2010})}\BibitemShut {NoStop}%
\bibitem [{\citenamefont {Steinhauer}(2014)}]{Steinhauer:2}%
  \BibitemOpen
  \bibfield  {author} {\bibinfo {author} {\bibfnamefont {J.}~\bibnamefont
  {Steinhauer}},\ }\bibfield  {title} {\enquote {\bibinfo {title} {Observation
  of self-amplifying {H}awking radiation in an analogue black-hole laser},}\
  }\href {\doibase 10.1038/nphys3104} {\bibfield  {journal} {\bibinfo
  {journal} {Nat. Phys.}\ }\textbf {\bibinfo {volume} {10}},\ \bibinfo {pages}
  {864--869} (\bibinfo {year} {2014})}\BibitemShut {NoStop}%
\bibitem [{\citenamefont {Jacobson}\ and\ \citenamefont
  {Volovik}(1998)}]{He-3-analog}%
  \BibitemOpen
  \bibfield  {author} {\bibinfo {author} {\bibfnamefont {T.~A.}\ \bibnamefont
  {Jacobson}}\ and\ \bibinfo {author} {\bibfnamefont {G.~E.}\ \bibnamefont
  {Volovik}},\ }\bibfield  {title} {\enquote {\bibinfo {title} {Event horizons
  and ergoregions in ${}^{3}\mathrm{He}$},}\ }\href {\doibase
  10.1103/PhysRevD.58.064021} {\bibfield  {journal} {\bibinfo  {journal} {Phys.
  Rev. D}\ }\textbf {\bibinfo {volume} {58}},\ \bibinfo {pages} {064021}
  (\bibinfo {year} {1998})}\BibitemShut {NoStop}%
\bibitem [{\citenamefont {Leonhardt}\ and\ \citenamefont
  {Piwnicki}(1999)}]{Slow-light:1}%
  \BibitemOpen
  \bibfield  {author} {\bibinfo {author} {\bibfnamefont {U.}~\bibnamefont
  {Leonhardt}}\ and\ \bibinfo {author} {\bibfnamefont {P.}~\bibnamefont
  {Piwnicki}},\ }\bibfield  {title} {\enquote {\bibinfo {title} {Optics of
  nonuniformly moving media},}\ }\href {\doibase 10.1103/PhysRevA.60.4301}
  {\bibfield  {journal} {\bibinfo  {journal} {Phys. Rev. A}\ }\textbf {\bibinfo
  {volume} {60}},\ \bibinfo {pages} {4301--4312} (\bibinfo {year}
  {1999})}\BibitemShut {NoStop}%
\bibitem [{\citenamefont {Leonhardt}\ and\ \citenamefont
  {Piwnicki}(2000{\natexlab{a}})}]{Slow-light:2}%
  \BibitemOpen
  \bibfield  {author} {\bibinfo {author} {\bibfnamefont {U.}~\bibnamefont
  {Leonhardt}}\ and\ \bibinfo {author} {\bibfnamefont {P.}~\bibnamefont
  {Piwnicki}},\ }\bibfield  {title} {\enquote {\bibinfo {title} {Relativistic
  {E}ffects of {L}ight in {M}oving {M}edia with {E}xtremely {L}ow {G}roup
  {V}elocity},}\ }\href {\doibase 10.1103/PhysRevLett.84.822} {\bibfield
  {journal} {\bibinfo  {journal} {Phys. Rev. Lett.}\ }\textbf {\bibinfo
  {volume} {84}},\ \bibinfo {pages} {822--825} (\bibinfo {year}
  {2000}{\natexlab{a}})}\BibitemShut {NoStop}%
\bibitem [{\citenamefont {Visser}(2000)}]{Slow-light:3}%
  \BibitemOpen
  \bibfield  {author} {\bibinfo {author} {\bibfnamefont {M.}~\bibnamefont
  {Visser}},\ }\bibfield  {title} {\enquote {\bibinfo {title} {Comment on
  `{R}elativistic {E}ffects of {L}ight in {M}oving {M}edia with {E}xtremely
  {L}ow {G}roup {V}elocity'},}\ }\href {\doibase 10.1103/PhysRevLett.85.5252}
  {\bibfield  {journal} {\bibinfo  {journal} {Phys. Rev. Lett.}\ }\textbf
  {\bibinfo {volume} {85}},\ \bibinfo {pages} {5252} (\bibinfo {year}
  {2000})}\BibitemShut {NoStop}%
\bibitem [{\citenamefont {Leonhardt}\ and\ \citenamefont
  {Piwnicki}(2000{\natexlab{b}})}]{Slow-light:4}%
  \BibitemOpen
  \bibfield  {author} {\bibinfo {author} {\bibfnamefont {U.}~\bibnamefont
  {Leonhardt}}\ and\ \bibinfo {author} {\bibfnamefont {P.}~\bibnamefont
  {Piwnicki}},\ }\bibfield  {title} {\enquote {\bibinfo {title} {Leonhardt and
  {P}iwnicki {R}eply:},}\ }\href {\doibase 10.1103/PhysRevLett.85.5253}
  {\bibfield  {journal} {\bibinfo  {journal} {Phys. Rev. Lett.}\ }\textbf
  {\bibinfo {volume} {85}},\ \bibinfo {pages} {5253} (\bibinfo {year}
  {2000}{\natexlab{b}})}\BibitemShut {NoStop}%
\bibitem [{\citenamefont {Unruh}\ and\ \citenamefont
  {Sch\"utzhold}(2003)}]{Slow-light:5}%
  \BibitemOpen
  \bibfield  {author} {\bibinfo {author} {\bibfnamefont {W.~G.}\ \bibnamefont
  {Unruh}}\ and\ \bibinfo {author} {\bibfnamefont {R.}~\bibnamefont
  {Sch\"utzhold}},\ }\bibfield  {title} {\enquote {\bibinfo {title} {On slow
  light as a black hole analogue},}\ }\href {\doibase
  10.1103/PhysRevD.68.024008} {\bibfield  {journal} {\bibinfo  {journal} {Phys.
  Rev. D}\ }\textbf {\bibinfo {volume} {68}},\ \bibinfo {pages} {024008}
  (\bibinfo {year} {2003})}\BibitemShut {NoStop}%
\bibitem [{\citenamefont {Sch\"utzhold}\ and\ \citenamefont
  {Unruh}(2002)}]{Schuetzhold+Unruh:Surface-waves-analog}%
  \BibitemOpen
  \bibfield  {author} {\bibinfo {author} {\bibfnamefont {R.}~\bibnamefont
  {Sch\"utzhold}}\ and\ \bibinfo {author} {\bibfnamefont {W.~G.}\ \bibnamefont
  {Unruh}},\ }\bibfield  {title} {\enquote {\bibinfo {title} {Gravity wave
  analogues of black holes},}\ }\href {\doibase 10.1103/PhysRevD.66.044019}
  {\bibfield  {journal} {\bibinfo  {journal} {Phys. Rev. D}\ }\textbf {\bibinfo
  {volume} {66}},\ \bibinfo {pages} {044019} (\bibinfo {year}
  {2002})}\BibitemShut {NoStop}%
\bibitem [{\citenamefont {Giovanazzi}(2005)}]{Fermi-gas-analog}%
  \BibitemOpen
  \bibfield  {author} {\bibinfo {author} {\bibfnamefont {S.}~\bibnamefont
  {Giovanazzi}},\ }\bibfield  {title} {\enquote {\bibinfo {title} {Hawking
  {R}adiation in {S}onic {B}lack {H}oles},}\ }\href {\doibase
  10.1103/PhysRevLett.94.061302} {\bibfield  {journal} {\bibinfo  {journal}
  {Phys. Rev. Lett.}\ }\textbf {\bibinfo {volume} {94}},\ \bibinfo {pages}
  {061302} (\bibinfo {year} {2005})}\BibitemShut {NoStop}%
\bibitem [{\citenamefont {Sch\"utzhold}\ and\ \citenamefont
  {Unruh}(2005)}]{Schuetzhold+Unruh:Waveguide-analog}%
  \BibitemOpen
  \bibfield  {author} {\bibinfo {author} {\bibfnamefont {R.}~\bibnamefont
  {Sch\"utzhold}}\ and\ \bibinfo {author} {\bibfnamefont {W.~G.}\ \bibnamefont
  {Unruh}},\ }\bibfield  {title} {\enquote {\bibinfo {title} {Hawking
  {R}adiation in an {E}lectromagnetic {W}aveguide?}}\ }\href {\doibase
  10.1103/PhysRevLett.95.031301} {\bibfield  {journal} {\bibinfo  {journal}
  {Phys. Rev. Lett.}\ }\textbf {\bibinfo {volume} {95}},\ \bibinfo {pages}
  {031301} (\bibinfo {year} {2005})}\BibitemShut {NoStop}%
\bibitem [{\citenamefont {Horstmann}\ \emph {et~al.}(2011)\citenamefont
  {Horstmann}, \citenamefont {Sch\"utzhold}, \citenamefont {Reznik},
  \citenamefont {Fagnocchi},\ and\ \citenamefont
  {Cirac}}]{Horstwald:Ion-ring-analog}%
  \BibitemOpen
  \bibfield  {author} {\bibinfo {author} {\bibfnamefont {B.}~\bibnamefont
  {Horstmann}}, \bibinfo {author} {\bibfnamefont {R.}~\bibnamefont
  {Sch\"utzhold}}, \bibinfo {author} {\bibfnamefont {B.}~\bibnamefont
  {Reznik}}, \bibinfo {author} {\bibfnamefont {S.}~\bibnamefont {Fagnocchi}}, \
  and\ \bibinfo {author} {\bibfnamefont {J.~I.}\ \bibnamefont {Cirac}},\
  }\bibfield  {title} {\enquote {\bibinfo {title} {Hawking radiation on an ion
  ring in the quantum regime},}\ }\href {\doibase
  10.1088/1367-2630/13/4/045008} {\bibfield  {journal} {\bibinfo  {journal}
  {New J. Phys.}\ }\textbf {\bibinfo {volume} {13}},\ \bibinfo {pages} {045008}
  (\bibinfo {year} {2011})}\BibitemShut {NoStop}%
\bibitem [{\citenamefont {Rousseaux}\ \emph {et~al.}(2008)\citenamefont
  {Rousseaux}, \citenamefont {Mathis}, \citenamefont {Ma\"{\i}ssa},
  \citenamefont {Philbin},\ and\ \citenamefont
  {Leonhardt}}]{Surface-waves:Experiment-1}%
  \BibitemOpen
  \bibfield  {author} {\bibinfo {author} {\bibfnamefont {G.}~\bibnamefont
  {Rousseaux}}, \bibinfo {author} {\bibfnamefont {C.}~\bibnamefont {Mathis}},
  \bibinfo {author} {\bibfnamefont {P.}~\bibnamefont {Ma\"{\i}ssa}}, \bibinfo
  {author} {\bibfnamefont {T.~G.}\ \bibnamefont {Philbin}}, \ and\ \bibinfo
  {author} {\bibfnamefont {U.}~\bibnamefont {Leonhardt}},\ }\bibfield  {title}
  {\enquote {\bibinfo {title} {Observation of negative-frequency waves in a
  water tank: a classical analogue to the {H}awking effect?}}\ }\href {\doibase
  10.1088/1367-2630/10/5/053015} {\bibfield  {journal} {\bibinfo  {journal}
  {New J. Phys.}\ }\textbf {\bibinfo {volume} {10}},\ \bibinfo {pages} {053015}
  (\bibinfo {year} {2008})}\BibitemShut {NoStop}%
\bibitem [{\citenamefont {Weinfurtner}\ \emph {et~al.}(2011)\citenamefont
  {Weinfurtner}, \citenamefont {Tedford}, \citenamefont {Penrice},
  \citenamefont {Unruh},\ and\ \citenamefont
  {Lawrence}}]{Surface-waves:Experiment-2}%
  \BibitemOpen
  \bibfield  {author} {\bibinfo {author} {\bibfnamefont {S.}~\bibnamefont
  {Weinfurtner}}, \bibinfo {author} {\bibfnamefont {E.~W.}\ \bibnamefont
  {Tedford}}, \bibinfo {author} {\bibfnamefont {M.~C.~J.}\ \bibnamefont
  {Penrice}}, \bibinfo {author} {\bibfnamefont {W.~G.}\ \bibnamefont {Unruh}},
  \ and\ \bibinfo {author} {\bibfnamefont {G.~A.}\ \bibnamefont {Lawrence}},\
  }\bibfield  {title} {\enquote {\bibinfo {title} {Measurement of {S}timulated
  {H}awking {E}mission in an {A}nalogue {S}ystem},}\ }\href {\doibase
  10.1103/PhysRevLett.106.021302} {\bibfield  {journal} {\bibinfo  {journal}
  {Phys. Rev. Lett.}\ }\textbf {\bibinfo {volume} {106}},\ \bibinfo {pages}
  {021302} (\bibinfo {year} {2011})}\BibitemShut {NoStop}%
\bibitem [{\citenamefont {Sch\"utzhold}\ \emph {et~al.}(2002)\citenamefont
  {Sch\"utzhold}, \citenamefont {Plunien},\ and\ \citenamefont
  {Soff}}]{Schuetzhold:Dielectric-analog}%
  \BibitemOpen
  \bibfield  {author} {\bibinfo {author} {\bibfnamefont {R.}~\bibnamefont
  {Sch\"utzhold}}, \bibinfo {author} {\bibfnamefont {G.}~\bibnamefont
  {Plunien}}, \ and\ \bibinfo {author} {\bibfnamefont {G.}~\bibnamefont
  {Soff}},\ }\bibfield  {title} {\enquote {\bibinfo {title} {Dielectric {B}lack
  {H}ole {A}nalogs},}\ }\href {\doibase 10.1103/PhysRevLett.88.061101}
  {\bibfield  {journal} {\bibinfo  {journal} {Phys. Rev. Lett.}\ }\textbf
  {\bibinfo {volume} {88}},\ \bibinfo {pages} {061101} (\bibinfo {year}
  {2002})}\BibitemShut {NoStop}%
\bibitem [{\citenamefont {Philbin}\ \emph {et~al.}(2008)\citenamefont
  {Philbin}, \citenamefont {Kuklewicz}, \citenamefont {Robertson},
  \citenamefont {Hill}, \citenamefont {K\"onig},\ and\ \citenamefont
  {Leonhardt}}]{Philbin:Dielectric-analog}%
  \BibitemOpen
  \bibfield  {author} {\bibinfo {author} {\bibfnamefont {T.~G.}\ \bibnamefont
  {Philbin}}, \bibinfo {author} {\bibfnamefont {C.}~\bibnamefont {Kuklewicz}},
  \bibinfo {author} {\bibfnamefont {S.}~\bibnamefont {Robertson}}, \bibinfo
  {author} {\bibfnamefont {S.}~\bibnamefont {Hill}}, \bibinfo {author}
  {\bibfnamefont {F.}~\bibnamefont {K\"onig}}, \ and\ \bibinfo {author}
  {\bibfnamefont {U.}~\bibnamefont {Leonhardt}},\ }\bibfield  {title} {\enquote
  {\bibinfo {title} {Fiber-{O}ptical {A}nalog of the {E}vent {H}orizon},}\
  }\href {\doibase 10.1126/science.1153625} {\bibfield  {journal} {\bibinfo
  {journal} {Science}\ }\textbf {\bibinfo {volume} {319}},\ \bibinfo {pages}
  {1367--1370} (\bibinfo {year} {2008})}\BibitemShut {NoStop}%
\bibitem [{\citenamefont {Faccio}\ \emph {et~al.}(2010)\citenamefont {Faccio},
  \citenamefont {Cacciatori}, \citenamefont {Gorini}, \citenamefont {Sala},
  \citenamefont {Averchi}, \citenamefont {Lotti}, \citenamefont {Kolesik},\
  and\ \citenamefont {Moloney}}]{Faccio:Dielectric-analog-1}%
  \BibitemOpen
  \bibfield  {author} {\bibinfo {author} {\bibfnamefont {D.}~\bibnamefont
  {Faccio}}, \bibinfo {author} {\bibfnamefont {S.}~\bibnamefont {Cacciatori}},
  \bibinfo {author} {\bibfnamefont {V.}~\bibnamefont {Gorini}}, \bibinfo
  {author} {\bibfnamefont {V.~G.}\ \bibnamefont {Sala}}, \bibinfo {author}
  {\bibfnamefont {A.}~\bibnamefont {Averchi}}, \bibinfo {author} {\bibfnamefont
  {A.}~\bibnamefont {Lotti}}, \bibinfo {author} {\bibfnamefont
  {M.}~\bibnamefont {Kolesik}}, \ and\ \bibinfo {author} {\bibfnamefont
  {J.~V.}\ \bibnamefont {Moloney}},\ }\bibfield  {title} {\enquote {\bibinfo
  {title} {Analogue gravity and ultrashort laser pulse filamentation},}\ }\href
  {\doibase 10.1209/0295-5075/89/34004} {\bibfield  {journal} {\bibinfo
  {journal} {EPL}\ }\textbf {\bibinfo {volume} {89}},\ \bibinfo {pages} {34004}
  (\bibinfo {year} {2010})}\BibitemShut {NoStop}%
\bibitem [{\citenamefont {Belgiorno}\ \emph
  {et~al.}(2011{\natexlab{a}})\citenamefont {Belgiorno}, \citenamefont
  {Cacciatori}, \citenamefont {Ortenzi}, \citenamefont {Rizzi}, \citenamefont
  {Gorini},\ and\ \citenamefont {Faccio}}]{Faccio:Dielectric-analog-2}%
  \BibitemOpen
  \bibfield  {author} {\bibinfo {author} {\bibfnamefont {F.}~\bibnamefont
  {Belgiorno}}, \bibinfo {author} {\bibfnamefont {S.~L.}\ \bibnamefont
  {Cacciatori}}, \bibinfo {author} {\bibfnamefont {G.}~\bibnamefont {Ortenzi}},
  \bibinfo {author} {\bibfnamefont {L.}~\bibnamefont {Rizzi}}, \bibinfo
  {author} {\bibfnamefont {V.}~\bibnamefont {Gorini}}, \ and\ \bibinfo {author}
  {\bibfnamefont {D.}~\bibnamefont {Faccio}},\ }\bibfield  {title} {\enquote
  {\bibinfo {title} {Dielectric black holes induced by a refractive index
  perturbation and the {H}awking effect},}\ }\href {\doibase
  10.1103/PhysRevD.83.024015} {\bibfield  {journal} {\bibinfo  {journal} {Phys.
  Rev. D}\ }\textbf {\bibinfo {volume} {83}},\ \bibinfo {pages} {024015}
  (\bibinfo {year} {2011}{\natexlab{a}})}\BibitemShut {NoStop}%
\bibitem [{\citenamefont {Finazzi}\ and\ \citenamefont
  {Carusotto}(2012)}]{Finazzi+Carusotto:Kinematic}%
  \BibitemOpen
  \bibfield  {author} {\bibinfo {author} {\bibfnamefont {S.}~\bibnamefont
  {Finazzi}}\ and\ \bibinfo {author} {\bibfnamefont {I.}~\bibnamefont
  {Carusotto}},\ }\bibfield  {title} {\enquote {\bibinfo {title} {Kinematic
  study of the effect of dispersion in quantum vacuum emission from strong
  laser pulses},}\ }\href {\doibase 10.1140/epjp/i2012-12078-x} {\bibfield
  {journal} {\bibinfo  {journal} {Eur. Phys. J. Plus}\ }\textbf {\bibinfo
  {volume} {127}},\ \bibinfo {pages} {78} (\bibinfo {year} {2012})}\BibitemShut
  {NoStop}%
\bibitem [{\citenamefont {Rubino}\ \emph {et~al.}(2012)\citenamefont {Rubino},
  \citenamefont {Lotti}, \citenamefont {Belgiorno}, \citenamefont {Cacciatori},
  \citenamefont {Couairon}, \citenamefont {Leonhardt},\ and\ \citenamefont
  {Faccio}}]{Rubino:Soliton-induced}%
  \BibitemOpen
  \bibfield  {author} {\bibinfo {author} {\bibfnamefont {E.}~\bibnamefont
  {Rubino}}, \bibinfo {author} {\bibfnamefont {A.}~\bibnamefont {Lotti}},
  \bibinfo {author} {\bibfnamefont {F.}~\bibnamefont {Belgiorno}}, \bibinfo
  {author} {\bibfnamefont {S.~L.}\ \bibnamefont {Cacciatori}}, \bibinfo
  {author} {\bibfnamefont {A.}~\bibnamefont {Couairon}}, \bibinfo {author}
  {\bibfnamefont {U.}~\bibnamefont {Leonhardt}}, \ and\ \bibinfo {author}
  {\bibfnamefont {D.}~\bibnamefont {Faccio}},\ }\bibfield  {title} {\enquote
  {\bibinfo {title} {Soliton-induced relativistic-scattering and
  amplification},}\ }\href {\doibase 10.1038/srep00932} {\bibfield  {journal}
  {\bibinfo  {journal} {Sci. Rep.}\ }\textbf {\bibinfo {volume} {2}},\ \bibinfo
  {pages} {932} (\bibinfo {year} {2012})}\BibitemShut {NoStop}%
\bibitem [{\citenamefont {Finazzi}\ and\ \citenamefont
  {Carusotto}(2013)}]{Finazzi:Hopfield-model-results}%
  \BibitemOpen
  \bibfield  {author} {\bibinfo {author} {\bibfnamefont {S.}~\bibnamefont
  {Finazzi}}\ and\ \bibinfo {author} {\bibfnamefont {I.}~\bibnamefont
  {Carusotto}},\ }\bibfield  {title} {\enquote {\bibinfo {title} {Quantum
  vacuum emission in a nonlinear optical medium illuminated by a strong laser
  pulse},}\ }\href {\doibase 10.1103/PhysRevA.87.023803} {\bibfield  {journal}
  {\bibinfo  {journal} {Phys. Rev. A}\ }\textbf {\bibinfo {volume} {87}},\
  \bibinfo {pages} {023803} (\bibinfo {year} {2013})}\BibitemShut {NoStop}%
\bibitem [{\citenamefont {Petev}\ \emph {et~al.}(2013)\citenamefont {Petev},
  \citenamefont {Westerberg}, \citenamefont {Moss}, \citenamefont {Rubino},
  \citenamefont {Rimoldi}, \citenamefont {Cacciatori}, \citenamefont
  {Belgiorno},\ and\ \citenamefont {Faccio}}]{Petev:Moving-medium}%
  \BibitemOpen
  \bibfield  {author} {\bibinfo {author} {\bibfnamefont {M.}~\bibnamefont
  {Petev}}, \bibinfo {author} {\bibfnamefont {N.}~\bibnamefont {Westerberg}},
  \bibinfo {author} {\bibfnamefont {D.}~\bibnamefont {Moss}}, \bibinfo {author}
  {\bibfnamefont {E.}~\bibnamefont {Rubino}}, \bibinfo {author} {\bibfnamefont
  {C.}~\bibnamefont {Rimoldi}}, \bibinfo {author} {\bibfnamefont {S.~L.}\
  \bibnamefont {Cacciatori}}, \bibinfo {author} {\bibfnamefont
  {F.}~\bibnamefont {Belgiorno}}, \ and\ \bibinfo {author} {\bibfnamefont
  {D.}~\bibnamefont {Faccio}},\ }\bibfield  {title} {\enquote {\bibinfo {title}
  {Blackbody {E}mission from {L}ight {I}nteracting with an {E}ffective {M}oving
  {D}ispersive {M}edium},}\ }\href {\doibase 10.1103/PhysRevLett.111.043902}
  {\bibfield  {journal} {\bibinfo  {journal} {Phys. Rev. Lett.}\ }\textbf
  {\bibinfo {volume} {111}},\ \bibinfo {pages} {043902} (\bibinfo {year}
  {2013})}\BibitemShut {NoStop}%
\bibitem [{\citenamefont {Belgiorno}\ \emph {et~al.}(2014)\citenamefont
  {Belgiorno}, \citenamefont {Cacciatori},\ and\ \citenamefont {{Dalla
  Piazza}}}]{Belgiorno:Perturbative-photon-production}%
  \BibitemOpen
  \bibfield  {author} {\bibinfo {author} {\bibfnamefont {F.}~\bibnamefont
  {Belgiorno}}, \bibinfo {author} {\bibfnamefont {S.~L.}\ \bibnamefont
  {Cacciatori}}, \ and\ \bibinfo {author} {\bibfnamefont {F.}~\bibnamefont
  {{Dalla Piazza}}},\ }\bibfield  {title} {\enquote {\bibinfo {title}
  {Perturbative photon production in a dispersive medium},}\ }\href {\doibase
  10.1140/epjd/e2014-40803-6} {\bibfield  {journal} {\bibinfo  {journal} {Eur.
  Phys. J. D}\ }\textbf {\bibinfo {volume} {68}},\ \bibinfo {pages} {134}
  (\bibinfo {year} {2014})}\BibitemShut {NoStop}%
\bibitem [{\citenamefont {Finazzi}\ and\ \citenamefont
  {Carusotto}(2014)}]{Finazzi:Hopfield-model-WH}%
  \BibitemOpen
  \bibfield  {author} {\bibinfo {author} {\bibfnamefont {S.}~\bibnamefont
  {Finazzi}}\ and\ \bibinfo {author} {\bibfnamefont {I.}~\bibnamefont
  {Carusotto}},\ }\bibfield  {title} {\enquote {\bibinfo {title} {Spontaneous
  quantum emission from analog white holes in a nonlinear optical medium},}\
  }\href {\doibase 10.1103/PhysRevA.89.053807} {\bibfield  {journal} {\bibinfo
  {journal} {Phys. Rev. A}\ }\textbf {\bibinfo {volume} {89}},\ \bibinfo
  {pages} {053807} (\bibinfo {year} {2014})}\BibitemShut {NoStop}%
\bibitem [{\citenamefont {Belgiorno}\ \emph {et~al.}(2015)\citenamefont
  {Belgiorno}, \citenamefont {Cacciatori},\ and\ \citenamefont {{Dalla
  Piazza}}}]{Belgiorno:Hopfield-model-results}%
  \BibitemOpen
  \bibfield  {author} {\bibinfo {author} {\bibfnamefont {F.}~\bibnamefont
  {Belgiorno}}, \bibinfo {author} {\bibfnamefont {S.~L.}\ \bibnamefont
  {Cacciatori}}, \ and\ \bibinfo {author} {\bibfnamefont {F.}~\bibnamefont
  {{Dalla Piazza}}},\ }\bibfield  {title} {\enquote {\bibinfo {title} {Hawking
  effect in dielectric media and the {H}opfield model},}\ }\href {\doibase
  10.1103/PhysRevD.91.124063} {\bibfield  {journal} {\bibinfo  {journal} {Phys.
  Rev. D}\ }\textbf {\bibinfo {volume} {91}},\ \bibinfo {pages} {124063}
  (\bibinfo {year} {2015})}\BibitemShut {NoStop}%
\bibitem [{\citenamefont {Belgiorno}\ \emph {et~al.}(2016)\citenamefont
  {Belgiorno}, \citenamefont {Cacciatori},\ and\ \citenamefont {{Dalla
  Piazza}}}]{Belgiorno:Hopfield-revisited}%
  \BibitemOpen
  \bibfield  {author} {\bibinfo {author} {\bibfnamefont {F.}~\bibnamefont
  {Belgiorno}}, \bibinfo {author} {\bibfnamefont {S.~L.}\ \bibnamefont
  {Cacciatori}}, \ and\ \bibinfo {author} {\bibfnamefont {F.}~\bibnamefont
  {{Dalla Piazza}}},\ }\bibfield  {title} {\enquote {\bibinfo {title} {The
  {H}opfield model revisited: covariance and quantization},}\ }\href {\doibase
  10.1088/0031-8949/91/1/015001} {\bibfield  {journal} {\bibinfo  {journal}
  {Phys. Scr.}\ }\textbf {\bibinfo {volume} {91}},\ \bibinfo {pages} {015001}
  (\bibinfo {year} {2016})}\BibitemShut {NoStop}%
\bibitem [{\citenamefont {Belgiorno}\ \emph {et~al.}(2010)\citenamefont
  {Belgiorno}, \citenamefont {Cacciatori}, \citenamefont {Clerici},
  \citenamefont {Gorini}, \citenamefont {Ortenzi}, \citenamefont {Rizzi},
  \citenamefont {Rubino}, \citenamefont {Sala},\ and\ \citenamefont
  {Faccio}}]{Belgiorno:Experiment-1}%
  \BibitemOpen
  \bibfield  {author} {\bibinfo {author} {\bibfnamefont {F.}~\bibnamefont
  {Belgiorno}}, \bibinfo {author} {\bibfnamefont {S.~L.}\ \bibnamefont
  {Cacciatori}}, \bibinfo {author} {\bibfnamefont {M.}~\bibnamefont {Clerici}},
  \bibinfo {author} {\bibfnamefont {V.}~\bibnamefont {Gorini}}, \bibinfo
  {author} {\bibfnamefont {G.}~\bibnamefont {Ortenzi}}, \bibinfo {author}
  {\bibfnamefont {L.}~\bibnamefont {Rizzi}}, \bibinfo {author} {\bibfnamefont
  {E.}~\bibnamefont {Rubino}}, \bibinfo {author} {\bibfnamefont {V.~G.}\
  \bibnamefont {Sala}}, \ and\ \bibinfo {author} {\bibfnamefont
  {D.}~\bibnamefont {Faccio}},\ }\bibfield  {title} {\enquote {\bibinfo {title}
  {Hawking {R}adiation from {U}ltrashort {L}aser {P}ulse {F}ilaments},}\ }\href
  {\doibase 10.1103/PhysRevLett.105.203901} {\bibfield  {journal} {\bibinfo
  {journal} {Phys. Rev. Lett.}\ }\textbf {\bibinfo {volume} {105}},\ \bibinfo
  {pages} {203901} (\bibinfo {year} {2010})}\BibitemShut {NoStop}%
\bibitem [{\citenamefont {Rubino}\ \emph {et~al.}(2011)\citenamefont {Rubino},
  \citenamefont {Belgiorno}, \citenamefont {Cacciatori}, \citenamefont
  {Clerici}, \citenamefont {Gorini}, \citenamefont {Ortenzi}, \citenamefont
  {Rizzi}, \citenamefont {Sala}, \citenamefont {Kolesik},\ and\ \citenamefont
  {Faccio}}]{Rubino:Experiment-2}%
  \BibitemOpen
  \bibfield  {author} {\bibinfo {author} {\bibfnamefont {E.}~\bibnamefont
  {Rubino}}, \bibinfo {author} {\bibfnamefont {F.}~\bibnamefont {Belgiorno}},
  \bibinfo {author} {\bibfnamefont {S.~L.}\ \bibnamefont {Cacciatori}},
  \bibinfo {author} {\bibfnamefont {M.}~\bibnamefont {Clerici}}, \bibinfo
  {author} {\bibfnamefont {V.}~\bibnamefont {Gorini}}, \bibinfo {author}
  {\bibfnamefont {G.}~\bibnamefont {Ortenzi}}, \bibinfo {author} {\bibfnamefont
  {L.}~\bibnamefont {Rizzi}}, \bibinfo {author} {\bibfnamefont {V.~G.}\
  \bibnamefont {Sala}}, \bibinfo {author} {\bibfnamefont {M.}~\bibnamefont
  {Kolesik}}, \ and\ \bibinfo {author} {\bibfnamefont {D.}~\bibnamefont
  {Faccio}},\ }\bibfield  {title} {\enquote {\bibinfo {title} {Experimental
  evidence of analogue {H}awking radiation from ultrashort laser pulse
  filaments},}\ }\href {\doibase 10.1088/1367-2630/13/8/085005} {\bibfield
  {journal} {\bibinfo  {journal} {New J. Phys.}\ }\textbf {\bibinfo {volume}
  {13}},\ \bibinfo {pages} {085005} (\bibinfo {year} {2011})}\BibitemShut
  {NoStop}%
\bibitem [{\citenamefont {Sch\"utzhold}\ and\ \citenamefont
  {Unruh}(2011)}]{Schuetzhold:Comment-Belgiorno-experiment}%
  \BibitemOpen
  \bibfield  {author} {\bibinfo {author} {\bibfnamefont {R.}~\bibnamefont
  {Sch\"utzhold}}\ and\ \bibinfo {author} {\bibfnamefont {W.~G.}\ \bibnamefont
  {Unruh}},\ }\bibfield  {title} {\enquote {\bibinfo {title} {Comment on
  `{H}awking {R}adiation from {U}ltrashort {L}aser {P}ulse {F}ilaments'},}\
  }\href {\doibase 10.1103/PhysRevLett.107.149401} {\bibfield  {journal}
  {\bibinfo  {journal} {Phys. Rev. Lett.}\ }\textbf {\bibinfo {volume} {107}},\
  \bibinfo {pages} {149401} (\bibinfo {year} {2011})}\BibitemShut {NoStop}%
\bibitem [{\citenamefont {Belgiorno}\ \emph
  {et~al.}(2011{\natexlab{b}})\citenamefont {Belgiorno}, \citenamefont
  {Cacciatori}, \citenamefont {Clerici}, \citenamefont {Gorini}, \citenamefont
  {Ortenzi}, \citenamefont {Rizzi}, \citenamefont {Rubino}, \citenamefont
  {Sala},\ and\ \citenamefont {Faccio}}]{Belgiorno:Experiment-reply}%
  \BibitemOpen
  \bibfield  {author} {\bibinfo {author} {\bibfnamefont {F.}~\bibnamefont
  {Belgiorno}}, \bibinfo {author} {\bibfnamefont {S.~L.}\ \bibnamefont
  {Cacciatori}}, \bibinfo {author} {\bibfnamefont {M.}~\bibnamefont {Clerici}},
  \bibinfo {author} {\bibfnamefont {V.}~\bibnamefont {Gorini}}, \bibinfo
  {author} {\bibfnamefont {G.}~\bibnamefont {Ortenzi}}, \bibinfo {author}
  {\bibfnamefont {L.}~\bibnamefont {Rizzi}}, \bibinfo {author} {\bibfnamefont
  {E.}~\bibnamefont {Rubino}}, \bibinfo {author} {\bibfnamefont {V.~G.}\
  \bibnamefont {Sala}}, \ and\ \bibinfo {author} {\bibfnamefont
  {D.}~\bibnamefont {Faccio}},\ }\bibfield  {title} {\enquote {\bibinfo {title}
  {Belgiorno \textit{et al.} {R}eply:},}\ }\href {\doibase
  10.1103/PhysRevLett.107.149402} {\bibfield  {journal} {\bibinfo  {journal}
  {Phys. Rev. Lett.}\ }\textbf {\bibinfo {volume} {107}},\ \bibinfo {pages}
  {149402} (\bibinfo {year} {2011}{\natexlab{b}})}\BibitemShut {NoStop}%
\bibitem [{\citenamefont {Liberati}\ \emph {et~al.}(2012)\citenamefont
  {Liberati}, \citenamefont {Prain},\ and\ \citenamefont
  {Visser}}]{Liberati:On-Belgiorno-Experiment}%
  \BibitemOpen
  \bibfield  {author} {\bibinfo {author} {\bibfnamefont {S.}~\bibnamefont
  {Liberati}}, \bibinfo {author} {\bibfnamefont {A.}~\bibnamefont {Prain}}, \
  and\ \bibinfo {author} {\bibfnamefont {M.}~\bibnamefont {Visser}},\
  }\bibfield  {title} {\enquote {\bibinfo {title} {Quantum vacuum radiation in
  optical glass},}\ }\href {\doibase 10.1103/PhysRevD.85.084014} {\bibfield
  {journal} {\bibinfo  {journal} {Phys. Rev. D}\ }\textbf {\bibinfo {volume}
  {85}},\ \bibinfo {pages} {084014} (\bibinfo {year} {2012})}\BibitemShut
  {NoStop}%
\bibitem [{\citenamefont {Unruh}\ and\ \citenamefont
  {Sch\"utzhold}(2012)}]{Unruh+Schuetzhold:On-Belgiorno-experiment}%
  \BibitemOpen
  \bibfield  {author} {\bibinfo {author} {\bibfnamefont {W.~G.}\ \bibnamefont
  {Unruh}}\ and\ \bibinfo {author} {\bibfnamefont {R.}~\bibnamefont
  {Sch\"utzhold}},\ }\bibfield  {title} {\enquote {\bibinfo {title} {Hawking
  radiation from `phase horizons' in laser filaments?}}\ }\href {\doibase
  10.1103/PhysRevD.86.064006} {\bibfield  {journal} {\bibinfo  {journal} {Phys.
  Rev. D}\ }\textbf {\bibinfo {volume} {86}},\ \bibinfo {pages} {064006}
  (\bibinfo {year} {2012})}\BibitemShut {NoStop}%
\bibitem [{\citenamefont {Linder}(2013)}]{Linder:Diploma-thesis}%
  \BibitemOpen
  \bibfield  {author} {\bibinfo {author} {\bibfnamefont {M.~F.}\ \bibnamefont
  {Linder}},\ }\emph {\bibinfo {title} {Hawking radiation in dispersive
  dielectric media}},\ \href@noop {} {\bibinfo {type} {diploma thesis}},\
  \bibinfo  {school} {Universit\"at Duisburg-Essen} (\bibinfo {year}
  {2013})\BibitemShut {NoStop}%
\bibitem [{\citenamefont {Hopfield}(1958)}]{Hopfield-model:1}%
  \BibitemOpen
  \bibfield  {author} {\bibinfo {author} {\bibfnamefont {J.~J.}\ \bibnamefont
  {Hopfield}},\ }\bibfield  {title} {\enquote {\bibinfo {title} {Theory of the
  {C}ontribution of {E}xcitons to the {C}omplex {D}ielectric {C}onstant of
  {C}rystals},}\ }\href {\doibase 10.1103/PhysRev.112.1555} {\bibfield
  {journal} {\bibinfo  {journal} {Phys. Rev.}\ }\textbf {\bibinfo {volume}
  {112}},\ \bibinfo {pages} {1555--1567} (\bibinfo {year} {1958})}\BibitemShut
  {NoStop}%
\bibitem [{\citenamefont {Huttner}\ and\ \citenamefont
  {Barnett}(1992)}]{Hopfield-model:2}%
  \BibitemOpen
  \bibfield  {author} {\bibinfo {author} {\bibfnamefont {B.}~\bibnamefont
  {Huttner}}\ and\ \bibinfo {author} {\bibfnamefont {S.~M.}\ \bibnamefont
  {Barnett}},\ }\bibfield  {title} {\enquote {\bibinfo {title} {Quantization of
  the electromagnetic field in dielectrics},}\ }\href {\doibase
  10.1103/PhysRevA.46.4306} {\bibfield  {journal} {\bibinfo  {journal} {Phys.
  Rev. A}\ }\textbf {\bibinfo {volume} {46}},\ \bibinfo {pages} {4306--4322}
  (\bibinfo {year} {1992})}\BibitemShut {NoStop}%
\bibitem [{\citenamefont {Faccio}\ \emph {et~al.}(2012)\citenamefont {Faccio},
  \citenamefont {Arane}, \citenamefont {Lamperti},\ and\ \citenamefont
  {Leonhardt}}]{Faccio:BH-lasers}%
  \BibitemOpen
  \bibfield  {author} {\bibinfo {author} {\bibfnamefont {D.}~\bibnamefont
  {Faccio}}, \bibinfo {author} {\bibfnamefont {T.}~\bibnamefont {Arane}},
  \bibinfo {author} {\bibfnamefont {M.}~\bibnamefont {Lamperti}}, \ and\
  \bibinfo {author} {\bibfnamefont {U.}~\bibnamefont {Leonhardt}},\ }\bibfield
  {title} {\enquote {\bibinfo {title} {Optical black hole lasers},}\ }\href
  {\doibase 10.1088/0264-9381/29/22/224009} {\bibfield  {journal} {\bibinfo
  {journal} {Class. Quantum Grav.}\ }\textbf {\bibinfo {volume} {29}},\
  \bibinfo {pages} {224009} (\bibinfo {year} {2012})}\BibitemShut {NoStop}%
\bibitem [{\citenamefont {Sch\"utzhold}\ and\ \citenamefont
  {Unruh}(2013)}]{Schuetzhold+Unruh:WKB-breakdown}%
  \BibitemOpen
  \bibfield  {author} {\bibinfo {author} {\bibfnamefont {R.}~\bibnamefont
  {Sch\"utzhold}}\ and\ \bibinfo {author} {\bibfnamefont {W.~G.}\ \bibnamefont
  {Unruh}},\ }\bibfield  {title} {\enquote {\bibinfo {title} {Hawking radiation
  with dispersion versus breakdown of the {WKB} approximation},}\ }\href
  {\doibase 10.1103/PhysRevD.88.124009} {\bibfield  {journal} {\bibinfo
  {journal} {Phys. Rev. D}\ }\textbf {\bibinfo {volume} {88}},\ \bibinfo
  {pages} {124009} (\bibinfo {year} {2013})}\BibitemShut {NoStop}%
\bibitem [{\citenamefont {Wald}(1984)}]{Wald:GR}%
  \BibitemOpen
  \bibfield  {author} {\bibinfo {author} {\bibfnamefont {R.~M.}\ \bibnamefont
  {Wald}},\ }\href@noop {} {\emph {\bibinfo {title} {{General Relativity}}}}\
  (\bibinfo  {publisher} {University of Chicago Press},\ \bibinfo {address}
  {Chicago},\ \bibinfo {year} {1984})\BibitemShut {NoStop}%
\bibitem [{\citenamefont {Wong}(2001)}]{Wong:Asymptotics}%
  \BibitemOpen
  \bibfield  {author} {\bibinfo {author} {\bibfnamefont {R.}~\bibnamefont
  {Wong}},\ }\href {\doibase 10.1137/1.9780898719260} {\emph {\bibinfo {title}
  {{Asymptotic Approximations of Integrals}}}}\ (\bibinfo  {publisher} {Society
  for Industrial and Applied Mathematics},\ \bibinfo {address} {Philadelphia},\
  \bibinfo {year} {2001})\BibitemShut {NoStop}%
\end{thebibliography}

\end{document}